\documentclass[%
 reprint,
 amsmath,amssymb,
prb,
]{revtex4-2}

\usepackage[subpreambles=true]{standalone}
\usepackage{import}
\usepackage{graphicx}
\usepackage{dcolumn}
\usepackage{bm}

\begin{document}


\title{Layer group classification of two-dimensional materials}

\author{Jingheng Fu}
    \affiliation{State Key Laboratory of Low Dimensional Quantum Physics and Department of Physics, Tsinghua University, Beijing, 100084, China}
\author{Mikael Kuisma}%

\affiliation{%
 CAMD, Computational Atomic-Scale Materials Design, Department of Physics, Technical University of Denmark, 2800, Kgs. Lyngby, Denmark}
 \author{Ask Hjorth Larsen}%
 
\affiliation{%
 CAMD, Computational Atomic-Scale Materials Design, Department of Physics, Technical University of Denmark, 2800, Kgs. Lyngby, Denmark}
\author{Kohei Shinohara}
    \affiliation{Department of Materials Science and Engineering, Kyoto University, Sakyo, Kyoto 606-8501, Japan}
\author{Atsushi Togo}%
    \affiliation{Center for Basic Research on Materials, National Institute for Materials Science, Tsukuba, Ibaraki 305-0047, Japan}
\author{Kristian S. Thygesen}%
 \email{thygesen@fysik.dtu.dk}
\affiliation{%
 CAMD, Computational Atomic-Scale Materials Design, Department of Physics, Technical University of Denmark, 2800, Kgs. Lyngby, Denmark}


\date{\today}

\begin{abstract}
The symmetry of a crystal structure with a three-dimensional (3D) lattice can be classified by one of the 230 space group types. For some types of crystals, e.g. crystalline films, surfaces, or planar interfaces, it is more appropriate to assume a 2D lattice. With this assumption, the structure can be classified by one of the 80 layer group types. We have implemented an algorithm to determine the layer group type of a 3D structure with a 2D lattice, and applied it to more than 15\,000 monolayer structures in the Computational 2D Materials (C2DB) database. We compare the classification of monolayers by layer groups and space groups, respectively. The latter is defined as the space group of the 3D bulk structure obtained by repeating the monolayer periodically in the direction perpendicular to the 2D lattice (AA-stacking). By this correspondence, nine pairs of layer group types are mapped to the same space group type due to the inability of the space group to distinguish the in-plane and out-of-plane axes. In total 18\% of the monolayers in the C2DB belong to one of these layer group pairs and are thus not properly classified by the space group type. 
\end{abstract}


\maketitle


\section{Introduction}
The classification of periodic solids by means of symmetry groups is fundamental to our understanding and description of crystalline materials. In addition to providing a unique classification scheme for crystal structures, the symmetry group can be used to derive important qualitative information about the physical properties of a material before they are measured or calculated. For example, the symmetry group dictates the possible degeneracy of energy levels, governs the selection rules for electronic and vibronic transitions~\cite{hammermesh1963group}, determines the possible topological phases~\cite{wieder2022topological}, and relates different spatial components of response tensors of crystals~\cite{boyd2020nonlinear}. 

The symmetry of a three-dimensional (3D) crystal structure with 3D translation periodicity, i.e. a 3D lattice, is described by one of the 230 space group types. For a 3D crystal structure with a 2D lattice, for example a crystalline film, the relevant symmetry group is represented by the layer group. There exist a total of 80 layer group types in contrast to the 230 space group types. For the point group operations of the layer group, the symmetry axis of the operations must be either parallel or perpendicular to the 2D lattice plane. Moreover, the direction of the translation part of the symmetry operations must be parallel to the lattice plane.

Materials with a 2D lattice and a thickness of only a few atoms are generally referred to as 2D materials. This class of materials has attracted enormous attention from a broad range of scientific communities due to their unique and easily tunable properties, their extreme thinness, and their good device integration potential~\cite{briggs2019roadmap,novoselov20162d,luo2016recent,peimyoo2021electrical,zhao2020engineering}. Despite of this interest, the symmetry characterization of 2D materials has not received the attention it deserves, and is often restricted to specification of the point group or the space group of some related layered bulk structures. This holds in particular for all the computational databases of 2D materials, e.g. the Computational 2D Materials Database (C2DB)~\cite{haastrup2018computational,gjerding2021recent}, the Materials Cloud\cite{mounet2018two}, and the 2DMatPedia\cite{zhou20192dmatpedia}, which hold the atomic structures of thousands of monolayers. Similar to the vast majority of studies on 2D materials, these databases have classified the monolayers by the space group of the bulk structure obtained by repeating the monolayer periodically in the direction normal to the 2D lattice plane, i.e. the space group of the AA-stacked bulk crystal. An important reason for this state of affairs has been the lack of a readily accessible software tool to determine the layer group of a general 2D atomic structure.  

Here we report on a numerical algorithm to determine the layer group of a 3D crystal structure with a 2D lattice. The algorithm has been implemented and made available as part of the open source space group and magnetic space group symmetry analysis library \texttt{Spglib}~\cite{spglibv1,shinohara2023algorithms}, which is widely used by material design~\cite{Avogadro,PyXtal,WulffPack}, simulation~\cite{CP2K,DFTKjcon,ICET,Octopus} and analysis~\cite{AiiDA,ASE,Matminer,Nexus,pymatgen,hiphive,Phonopy,SeeKpath} programs and various high-throughput material databases~\cite{JARVIS,MaterialsProject,OQMD,haastrup2018computational,gjerding2021recent,mounet2018two,zhou20192dmatpedia}. We use the algorithm to obtain the layer groups of the 15733 crystal structures currently contained in the C2DB database, and we present a statistical analysis of the data. This analysis reveals a rather inhomogeneous distribution with a number of layer group types being either very scarcely represented or not represented at all. We also show that the classification of a monolayer by the space group of its AA-stacked bulk structure is incomplete. In fact, for 18\% of the monolayers in C2DB, the space group of the AA-stacked bulk crystal does not uniquely determine the layer group.

\section{Method}

In this paper, we use \textit{layer system} to refer to the collection of a layer group crystallographic coordinate system and the atomic structure defined in this coordinate system. Following Refs.~\onlinecite{spglibv1} and \onlinecite{ITE}, $\mathbf{a}$, $\mathbf{b}$ and $\mathbf{c}$ represent basis vectors of the system, with two of the basis vectors being lattice vectors. Without loss of generality, we take the lattice vectors to be $\mathbf{a}$ and $\mathbf{b}$. Furthermore, we denote the basis vectors in different unit cells by $(\mathbf{a}_{\mathrm{i}}, \mathbf{b}_{\mathrm{i}}, \mathbf{c}_{\mathrm{i}})$, $(\mathbf{a}_{\mathrm{p}}, \mathbf{b}_{\mathrm{p}}, \mathbf{c}_{\mathrm{p}})$, and $(\mathbf{a}_{\mathrm{c}}, \mathbf{b}_{\mathrm{c}}, \mathbf{c}_{\mathrm{c}})$, where the subscripts refer to the input cell, primitive cell, and conventional cell, respectively. Atom coordinates in the unit cell basis, i.e. fractional coordinates, are denoted by 
$\bm{x}$. When necessary, the subscript $\mathrm{i}$, $\mathrm{p}$ or $\mathrm{c}$ is used to indicate the type of unit cell. For layer groups, the role of $\mathbf{c}$ is only to provide a basis vector for representing the atomic positions. Nevertheless, the current implementation follows the crystallographic space group convention which requires the rotation part of symmetry transformations to be represented by integral matrices under primitive cell basis (see Sec. \ref{sec:symmetry}). Hence the choice of $\mathbf{c}$-basis can affect the point group and thus layer group. In this paper, we fix the non-lattice vector $\mathbf{c}_{\mathrm{i}}$ to be orthogonal to $\mathbf{a}_{\mathrm{i}}$ and $\mathbf{b}_{\mathrm{i}}$ to obtain the highest symmetry.

For crystals with periodicity along the three basis axes, the distance between two atoms with positions $\bm{x}$ and $\bm{x}'$ respectively is defined by 
\begin{equation}
\label{eq:red-Euclidean-dist}
d(\bm{x},\bm{x}')=\lVert (\mathbf{a}, \mathbf{b}, \mathbf{c}) (\Delta\bm{x} - \lfloor\Delta\bm{x}\rceil) \rVert .
\end{equation}
Here $(\mathbf{a}, \mathbf{b}, \mathbf{c})$ denotes the coordinate matrix of the unit cell basis vectors, $\Delta\bm{x} = \bm{x} - \bm{x}'$, and $\lfloor\cdot\rceil$ rounds components of a vector to the nearest integer. Layer systems do not have periodicity in the direction along the non-lattice vector $\mathbf{c}$, and the rounding operation in Eq.~(\ref{eq:red-Euclidean-dist}) is not applied to this coordinate. To account for small distortions or numerical noise, \texttt{Spglib} determines the equality of two atom positions by the condition 
\begin{equation}
\label{eq:pos-equal}
d(\bm{x},\bm{x}') < \epsilon,
\end{equation}
where $\epsilon$ is a small tolerance parameter (called `symprec') set by the user.

A symmetry operation maps the crystal structure onto itself. The operation is denoted $(\bm{W}, \bm{w})$, where the linear part $\bm{W}$ is a $3\times 3$ integral matrix, and the translation part $\bm{w}$ is a $3\times 1$ column vector. By the symmetry operation, an atom with position $\bm{x}$ will be sent to 
\begin{equation}
\tilde{\bm{x}} = \bm{W}\bm{x} + \bm{w}.
\end{equation}
Symmetry requires the position $\bm{x}'$ of one and only one atom with the same species to be equal to $\tilde{\bm{x}}$ in the sense of Eq.~(\ref{eq:pos-equal}).

\begin{figure*}
    \centering
    \includegraphics[width=\textwidth]{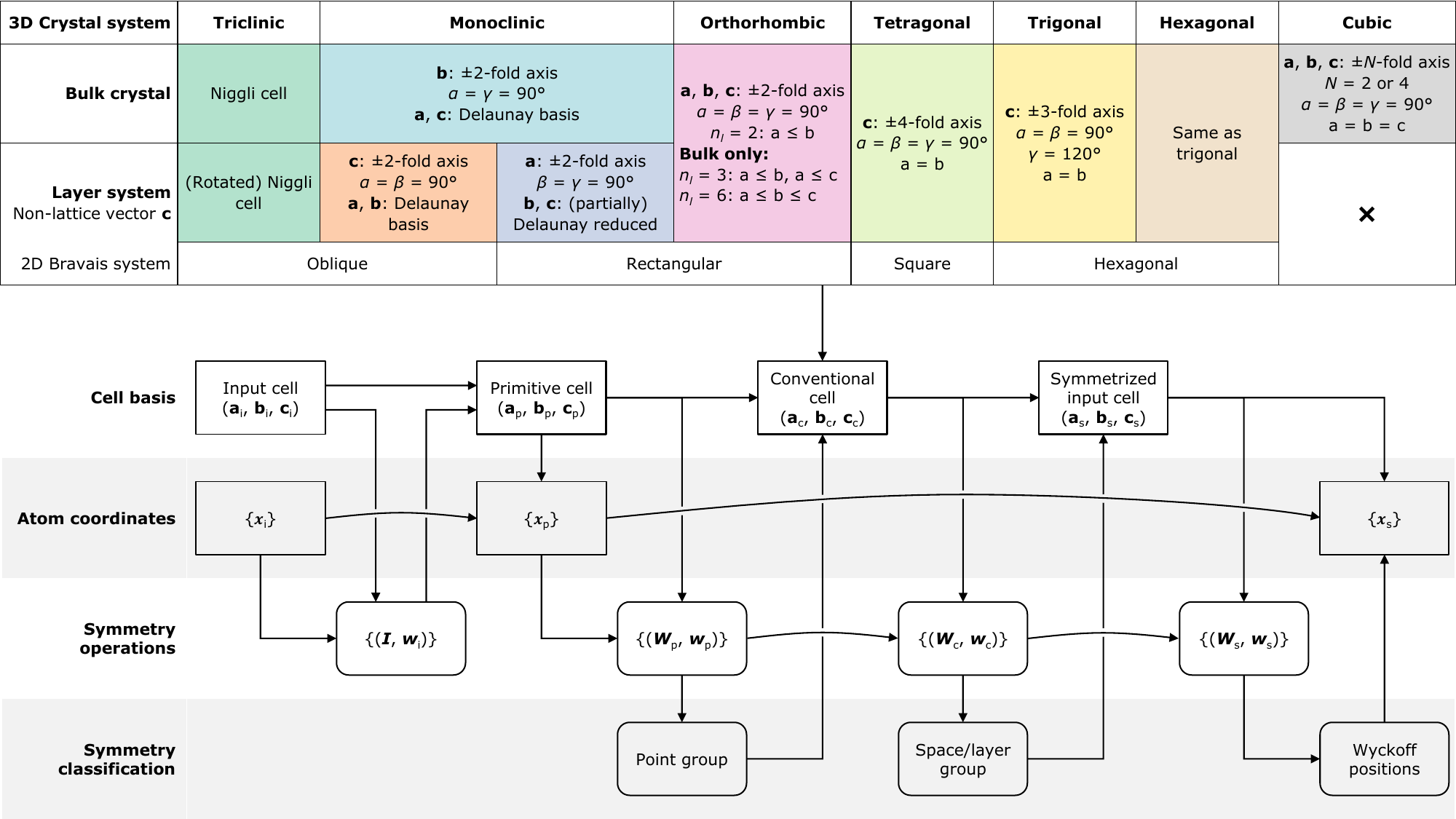}
    \caption{Upper: The criteria (rotation axes and metric conditions) used by \texttt{Spglib} to determine a standardized conventional cell of a bulk crystal (3D lattice) and a layer system (2D lattice). The criteria are listed for each of the seven crystal systems. The layer system can be additionally classified by the Bravais system of its lattice, which is listed in the last row of the table. For triclinic and monoclinic crystal systems, the criteria specific to bulk and layer systems are listed in the second and third rows, respectively. Monoclinic crystal systems with oblique and rectangular lattices have different criteria, and are separately listed with different colors. For orthorhombic systems, the bulk and layer systems share most of the conditions. For the tetragonal, trigonal and hexagonal crystal systems, the bulk and layer systems share all the same criteria, and the rows for these systems are combined. The cubic crystal system does not exist for layer groups, and is left blank in the table. The criteria include axis choices for selecting $\mathbf{a}_\mathrm{c}$, $\mathbf{b}_\mathrm{c}$ and $\mathbf{c}_\mathrm{c}$ from the symmetry operation with certain types, and metric parameters. The remaining conditions are additional metric rules used by \texttt{Spglib} when the standard choices defined in Refs.~\onlinecite{ITA} and \onlinecite{ITE} cannot uniquely determine the shape and/or setting of the conventional cells. Details of the rules are described in Appendix.~\ref{sec:conv-cell}. Lower: Workflow used by the algorithm to determine the point group, space/layer group, Wyckoff positions, and more.}
    \label{fig:workflow-conv-cell}
\end{figure*}

Our algorithm for determining the layer group of a layer system is illustrated in Fig.~\ref{fig:workflow-conv-cell}. The algorithm follows the one already implemented in \texttt{Spglib} for space groups and documented elsewhere~\cite{spglibv1}; thus we mainly focus on the aspects that are specific to layer groups. 
The lower part of Fig.~\ref{fig:workflow-conv-cell} shows the main steps required to determine the layer/space group while the upper part details the criteria used to select the conventional unit cell for the various crystal systems. In addition to determining the point group and space/layer group, the algorithm also determines the site symmetry group for each atom and the occupied Wyckoff positions, and symmetrizes the input structure.


\subsection{Primitive cell}
The input cell will be transformed to a provisional primitive cell to reduce the computational complexity. To determine the primitive cell, the first step is to check whether the input cell can be reduced to a smaller unit that generates the crystal. Specifically, the algorithm searches for symmetry operations that are pure translations, i.e. $\bm{W}_{\mathrm{i}}=\bm{I}$ and $\bm{w}_{\mathrm{i}} \neq 0$. The three vectors among the found $\bm{w}$s and the input cell basis $\mathbf{a}_\mathrm{i}$, $\mathbf{b}_\mathrm{i}$ and $\mathbf{c}_\mathrm{i}$ which form the cell with smallest non-zero volume are the primitive cell basis vectors. For layer systems, the non-lattice vector $\mathbf{c}_\mathrm{i}$ will always be chosen as the primitive basis vector $\mathbf{c}_\mathrm{p}$. The obtained primitive cell will further be transformed to the Delaunay cell~\cite{Delaunay1933}. The shape of the Delaunay primitive cell belongs to one of the 24 symmetric varieties~\cite{Delaunay1933, BurzlaffZimmermann1985}. Matrix elements of the linear part $\bm{W}_{\mathrm{p}}$ are 0, 1 or -1 under the Delaunay reduced basis. Note that this procedure may not be fully applicable to a layer system as it can transform the lattice vectors into linear combinations of the original lattice and non-lattice vectors. Thus the Delaunay reduction requires special care for layer systems as discussed in detail in Appendix.~\ref{sec:delaunay}.

\subsection{Symmetry operations}\label{sec:symmetry}
After obtaining the primitive cell, the algorithm searches for the matrix-column pairs $(\bm{W}_{\mathrm{p}}, \bm{w}_{\mathrm{p}})$ of all symmetry operations in the Delaunay reduced primitive cell basis $(\mathbf{a}_\mathrm{p}, \mathbf{b}_\mathrm{p}, \mathbf{c}_\mathrm{p})$. For space group operations, $\bm{W}_{\mathrm{p}}$ consists of 0 and $\pm 1$ satisfying $\det(\bm{W}_{\mathrm{p}}) = \pm 1$. For layer groups, preservation of the 2D lattice prohibits mixing the non-periodic component into the lattice vectors, e.g. $\mathbf{a}' = \mathbf{a} + \mathbf{c}$. For layer systems, `A'-face, `B'-face, body-, and face centred conventional cells do not exist. Consequently, these conventional cells are not considered. The algorithm searches for operations with the matrix pattern
\begin{equation}
\label{eq:layer-W-pattern}
\bm{W}_{\mathrm{layer}} = \begin{pmatrix}
                                W_{11} & W_{12} & 0 \\
                                W_{21} & W_{22} & 0 \\
                                     0 &      0 & \pm 1
                             \end{pmatrix}
\quad W_{ij} = 0, \pm 1.
\end{equation}
With this matrix pattern, any operation with rotation order higher than two must have the rotation axis perpendicular to the lattice plane. Consequently, point groups belonging to the cubic crystal system are not allowed for layer systems. In addition, any $\pm 2$-fold rotations must have the rotation axis either parallel or perpendicular to the lattice plane. Furthermore, Eq.~(\ref{eq:red-Euclidean-dist}) rules out the possibility of finding screw rotation or glide reflection with a non-lattice component in the translation part $\bm{w}$. To find the symmetry operations of a layer system, the algorithm searches for all transformations $(\bm{W}, \bm{w})$ with $\bm{W}$ of the form (\ref{eq:layer-W-pattern}) satisfying 
$d(\tilde{\bm{x}} , \bm{x}')<\epsilon$, where $d(\cdot,\cdot)$ is defined in Eq. (\ref{eq:red-Euclidean-dist}),  $\tilde{\bm{x}}=\bm{W}\bm{x}+\bm{w}$, and $\bm{x}$ and $\bm{x}'$ are the position of atoms of the same type.

\subsection{Point group}
The linear part $\bm{W}$ of all found symmetry operations $(\bm{W},\bm{w})$ compose the crystallographic point group $\mathcal{P}$ in the primitive cell basis. The type of the point group is determined by counting the number of the ten different rotation orders $(\pm1, \pm2, \pm3, \pm4, \pm6)$ appearing in $\mathcal{P}$ (see Table V in Ref.~\onlinecite{spglibv1}). Once the point group type has been determined, the search for the space/layer group can be narrowed down accordingly. 

\subsection{Standardized conventional cell}
To determine the space/layer group type from the symmetry operations they must be compared with those already stored in \texttt{Spglib}. However, a direct comparison requires that the same crystallographic coordinate systems (CCS), i.e. origin position and basis vectors (including their order), are employed. For each space group, Ref.~\onlinecite{ITA} lists the possible conventional CCSs and designates one of them as the standard choice. A standard CCS for layer systems is defined by Ref.~\onlinecite{ITE}.

\texttt{Spglib} adopts Hall symbols~\cite{Hall:a19707} to distinguish the different choices of CCS. A table of all the Hall symbols for space groups can be found in table A1.4.2.7 of Ref.~\onlinecite{ITB}. \texttt{Spglib} contains a database with all $(\bm{W}, \bm{w})$ matrix-column pairs generated using Hall symbols. For layer groups, we are not aware of a complete table of Hall symbols. Thus we have produced one corresponding to the different choices of CCSs listed in Ref.~\onlinecite{ITE}, see Appendix.~\ref{sec:hall-symbols}.  

\subsection{Layer group}
The crystallographic point group determines the type of the Bravais lattice and puts some constraints on the metric parameters of the conventional cell (relations between basis vector lengths and angles), see upper panel of Fig.~\ref{fig:workflow-conv-cell} and Appendix.~\ref{sec:conv-cell}. Generally, one of the basis vectors $\mathbf{a}_\mathrm{c}$, $\mathbf{b}_\mathrm{c}$ and $\mathbf{c}_\mathrm{c}$ is chosen as the symmetry axes of a specific symmetry operation, as shown in the first line for each criterion in Fig.~\ref{fig:workflow-conv-cell}. The shortest vectors in the perpendicular plane that satisfy the metric conditions are chosen for the remaining two basis vectors. The $\bm{W}$ matrices will retain integer elements under this basis.

When the basis vector choices are the same as the standard CCS of the relevant space/layer group, $\bm{W}$ will match the standard counterparts stored in \texttt{Spglib}. Still, $\bm{w}$ may differ from the stored ones by a shift of the origin. The origin shift is determined following the algorithm in Ref.~\onlinecite{spglibv1} and Ref.~\onlinecite{Grosse-Kunstleve:au0146} except the distances in the algorithm are measured by Eq.~(\ref{eq:red-Euclidean-dist}).


\subsection{Symmetrization}
The site symmetry group of each atom is determined using the distance measure in Eq.~(\ref{eq:red-Euclidean-dist}) and (\ref{eq:pos-equal}). Each atom is further assigned a Wyckoff position. Finally, the atomic structure is symmetrized by applying all of the site symmetry operations separately to each atom and performing an average over the resulting positions.
The standardized conventional cell is symmetrized such as to satisfy the lattice metric conditions in Fig.~\ref{fig:workflow-conv-cell}(upper) exactly to obtain the symmetrized input cell $(\mathbf{a}_\mathrm{s},\mathbf{b}_\mathrm{s},\mathbf{c}_\mathrm{s})$ and the standard primitive cell. For the standard conventional cells with primitive lattice, the standard primitive cells are the conventional cells. While for `C'-face centred standard conventional cells, the standard primitive cell $(\mathbf{a}_\mathrm{sp},\mathbf{b}_\mathrm{sp},\mathbf{c}_\mathrm{sp}) = (\mathbf{a}_\mathrm{sc},\mathbf{b}_\mathrm{sc},\mathbf{c}_\mathrm{sc}) \bm{P}_{\mathrm{C}}$, where
\begin{equation}
\label{eq:C-centring-mat}
\bm{P}_{\mathrm{C}} =
\renewcommand*{\arraystretch}{1.3}
\begin{pmatrix}
     \frac{1}{2}  & \frac{1}{2} & 0 \\
\bar{\frac{1}{2}} & \frac{1}{2} & 0 \\
                0 &           0 & 1
\end{pmatrix}.
\end{equation}


\begin{figure*}
\includegraphics[width=\textwidth]{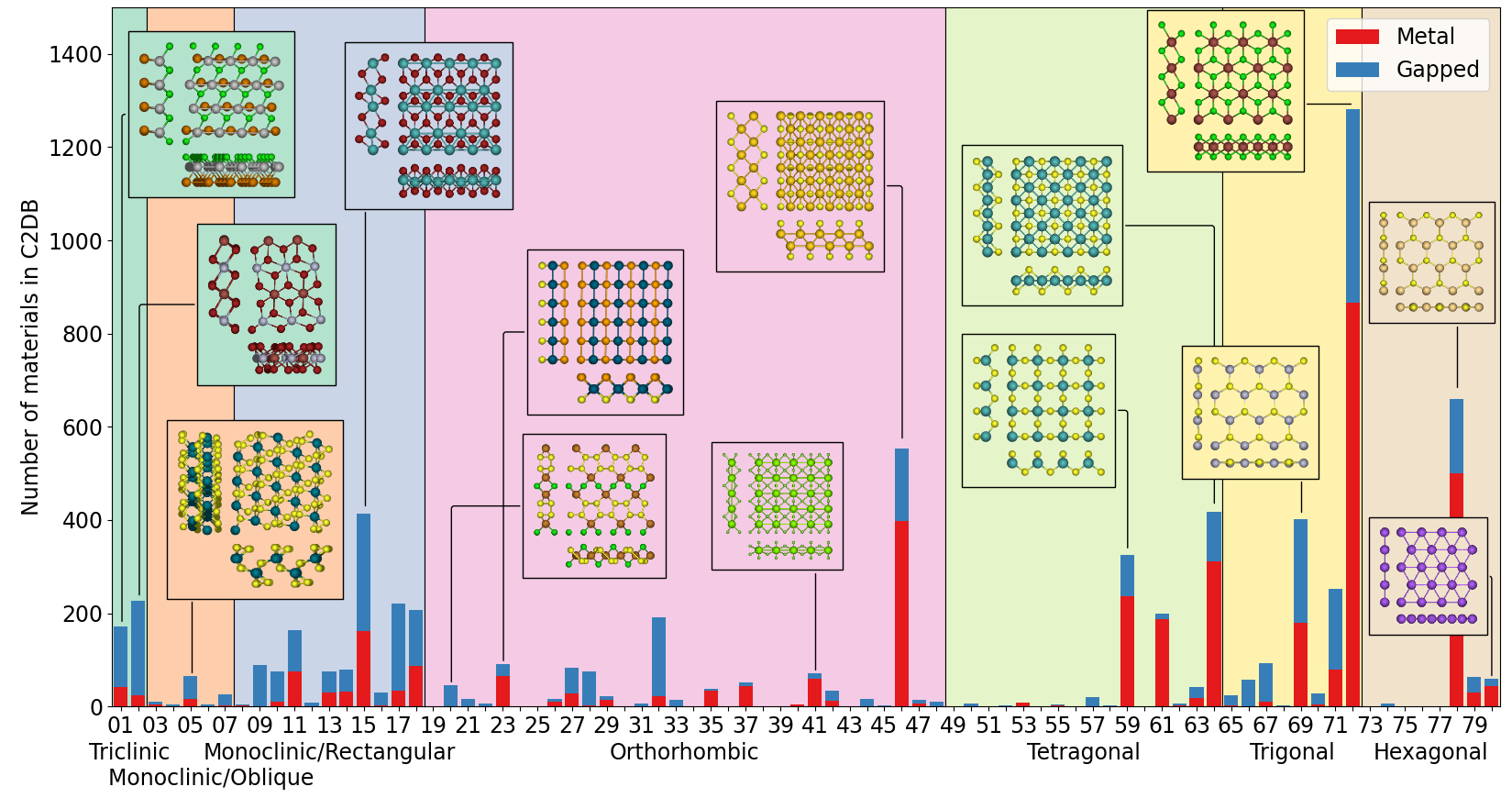}
\caption{Distribution of materials of C2DB according to layer group number. The background color coding refers to the crystal system (the monoclinic crystal system is further classified according to its lattice type). Some examples of structures from different layer groups are shown. The blue (red) part of the bars refers to materials with finite (zero) band gap.}
\label{MaterialDistribution}
\end{figure*}

\section{Results}
The Computational 2D Materials Database (C2DB) currently contains 15\,735 2D materials. Most of the structures have been generated by computational exfoliation of experimentally known layered bulk crystals followed by systematic lattice decoration (atomic substitution) of the thus obtained monolayers~\cite{haastrup2018computational,gjerding2021recent}. Recently, the data set was amended by monolayers created by a deep generative model~\cite{Lyngby2022}. All the monolayers in C2DB have been relaxed using density functional theory with the PBE exchange-correlation functional.   

We have determined the layer group of all the monolayers in C2DB alongside the space group of the corresponding AA-stacked bulk structures (see next section). Fig.~\ref{MaterialDistribution} shows the distribution of the C2DB monolayers according to the layer group. The distribution shown has been limited to the subset of monolayers for which the band gap has been calculated (7\,201 materials), and the materials have been divided into metallic and non-metallic compounds. The background color is used to indicate the different crystal systems (along with the Bravais system if necessary). Examples of crystal structures from selected layer groups are shown. 

The most frequently occurring layer group is number 72 ($p\bar{3}m1$) with 1\,231 occurrences. It should be noted, that the distribution of materials by layer group number is highly non-uniform, with several completely empty bins accompanied by several with just a few materials. In particular, there are no materials in C2DB with layer groups 19, 24, 25, 30, 38, 39, 43, 49, 54, 56, 60, 73, 75, 76 and 77 (using a tolerance parameter `symprec'=0.1\,\AA).

There is a tendency that insulating behaviour (blue bars) is more pronounced among the materials with lower layer group numbers while metallic behaviour (red bars) is dominant in materials with high layer group numbers. This is due to the fact that, as a rough rule, layer groups with smaller numbers contain fewer symmetry transformations. The less symmetric crystals will have energy bands of lower degeneracy and therefore a higher chance of all bands being either fully occupied or empty.
On the other hand, more symmetric crystals will have energy bands of higher degeneracy and consequently a higher chance of partially occupied, i.e. metallic, bands.

\begin{figure}
    \centering
    \includegraphics[width=2.4in]{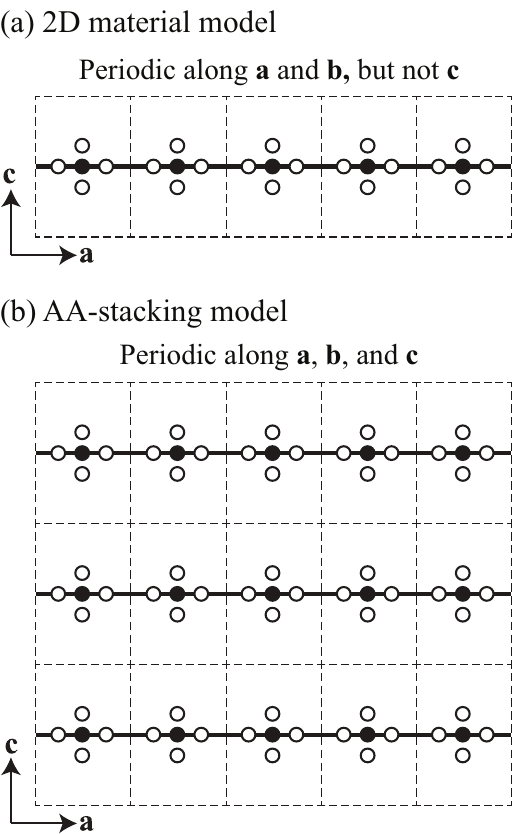}
    \caption{(a) The 2D material is periodic along the $\mathbf a$ and $\mathbf b$ axes, but not along the $\mathbf c$ axis. (b) In the AA-stacking model, the 2D material is periodically repeated along the out-of-plane $\mathbf c$ axis, extending it to the domain of space groups and making it compatible with computational plane wave codes, which require periodic boundary conditions in all directions.}
    \label{fig:AAstacking}
\end{figure}

\begin{figure}
\includegraphics[width=2.7in]{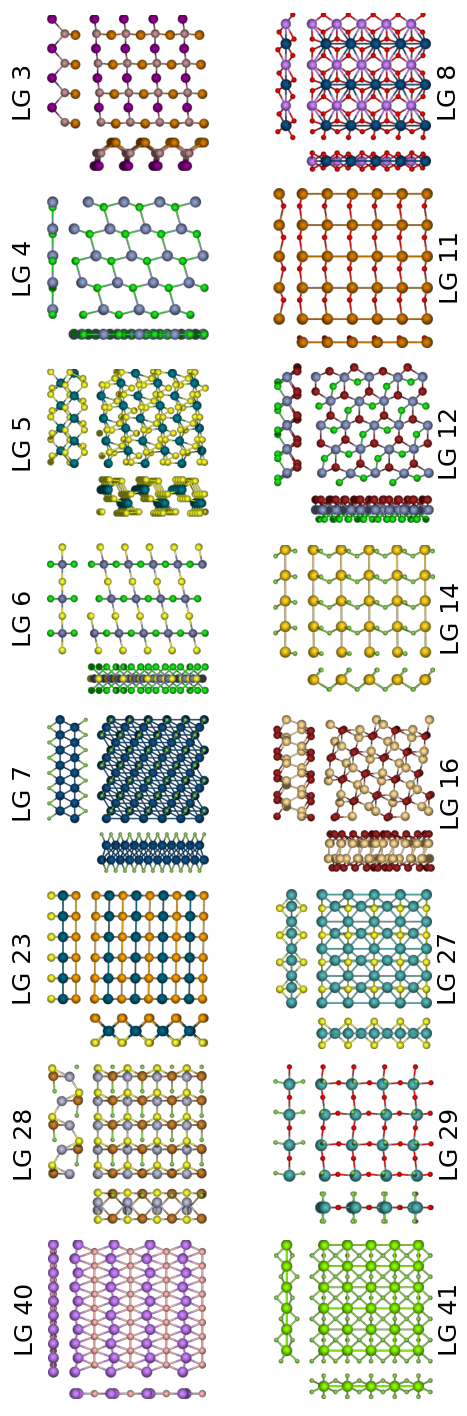}
\caption{Examples of structures that are incompletely classified by the space group of their AA-stacked bulk, cf. Table~\ref{tab:spgmapping}. Each row shows a pair of 2D crystal structures belonging to distinct layer group types, but with the same space group type (of the AA-stacked bulk). Structures from the layer group pair (24, 31) are not shown as the C2DB contains no monolayers from layer group 24.}
\label{fig:comparison}
\end{figure}

\subsection{Layer group versus space group}

As mentioned in the introduction, 2D materials have so far mainly been classified using space groups. This is true in particular for the C2DB, which has used the space group of the AA-stacked bulk materials to represent the crystal symmetry of a monolayer. The `2D material model' and the `AA-stacking model' are illustrated in Fig.~\ref{fig:AAstacking}. In this section, we compare the two crystal models and highlight possible pitfalls arising when using the AA-stacking model to describe a monolayer. 

First, on a general note, it has been shown that for each layer group $\mathrm{L}$ with point group $\mathcal{P}$ and translation lattice $\mathbf{L}_\mathrm{L}$, there exists a symmorphic representative which is the 3D space group $\mathrm{S}$ that shares the same point group and symmetry diagram~\cite{Kopsk&yacute;:bx0374, Kopsk&yacute;:bx0469}. The lattice of the symmorphic representative
\begin{equation}
\label{eq:lattice-division}
    \mathbf{L}_\mathrm{S} = \mathbf{L}_\mathrm{L} \oplus \mathbf{L}_\mathrm{R}
\end{equation}
is the direct sum of the 2D lattice of the layer group $\mathrm{L}$ and the 1D lattice $\mathbf{L}_\mathrm{R}$ of a symmorphic rod group $\mathrm{R}$ with point group $\mathcal{P}$.
The layer group and its symmorphic representative can be represented by the same Hermann-Mauguin symbol which only differs by the upper and lowercase of the first letter~\cite{Kopsk&yacute;:bx0469}. However, for some types of space groups, the manner of splitting the lattice $\mathbf{L}_\mathrm{S}$ into two $\mathcal{P}$-invariant sub-lattice is not unique. Consequently, one symmorphic representative may correspond to multiple layer groups with different types. Ambiguity thereupon arises.

To be more specific, the symmorphic representatives of the 80 layer group types belong to 71 different space group types. 62 of these 71 space group types have a one-to-one mapping with the corresponding layer group types, while the rest 9 space group types map to 2 different layer group types, respectively. In Table~\ref{tab:spgmapping}, we list these 9 non-injective mappings. The numbers in parenthesis represent the number of materials in C2DB with that layer group type. Examples of concrete structures from the relevant layer group types are provided in Fig.~\ref{fig:comparison}. Note that only eight cases are shown in Fig.~\ref{fig:comparison} as C2DB currently does not include any structures from layer group 24. As illustrated in Table~\ref{tab:spgmapping}, a total of 2\,848 of the monolayers in C2DB (18\%) belong to one of the layer group types in Table~\ref{tab:spgmapping}, and thus are not uniquely classified by the space group of the AA-stacked structure.

Inspection of the structures/layer groups in Fig.~\ref{fig:comparison} and Table \ref{tab:spgmapping} shows that the first five pairs are examples where the two layer groups belong to the monoclinic crystal systems with oblique and rectangular Bravais systems, respectively. In these cases, one of the structures has an in-plane $\pm2$-fold symmetry, which is replaced by an out-of-plane $\pm2$-fold symmetry in the other structure. The last four pairs of layer groups all belong to the orthorhombic crystal system. For each pair of the examples, the out-of-plane rotation, mirror reflection or glide translation is exchanged with an in-plane symmetry that has a different type. For example, the out-of-plane 2-fold rotation and one of the in-plane mirror reflections of layer group 23 ($pmm2$) becomes an out-of-plane mirror reflection and in-plane 2-fold rotation for layer group 27 ($pm2m$). These distinctions cannot be made by the space group of the AA-stacked bulk.

The ability of the layer groups to differentiate structures of the same space group type (defined via the AA-stacking model) is not without practical value. For example, Ji \textit{et al.} showed how layer groups can be used as a tool to determine possible outcomes of ferroelectricity when two monolayers are stacked to form a bilayer~\cite{PhysRevLett.130.146801}. They further divide the layer groups into distinct polar types, according to whether polarization is allowed in-plane or out-of-plane or both. An example is given by Fig.~\ref{fig:comparison}, where layer group 4 ($p11m$) is polar in-plane and layer group 11 ($pm11$) is polar out-of-plane, while both share the space group 6 ($Pm$).

\begin{table}
\centering
\begin{tabular}{l|l|l}
\hline
\hline
\multicolumn{2}{c|}{Layer group pairs} & Space group  \\
\hline
3: $p112$ (46) &  8: $p211$ (11) &   3: $P2$ \\
4: $p11m$ (59) &  11: $pm11$ (591) &   6: $Pm$ \\
5: $p11a$ (201) &  12: $pb11$ (11) &   7: $Pc$ \\
6: $p112/m$ (30) &  14: $p2/m11$ (230) &   10: $P2/m$ \\
7: $p112/a$ (51) &  16: $p2/b11$ (88) &   13: $P2/c$ \\
23: $pmm2$ (375) &  27: $pm2m$ (186) &   25: $Pmm2$ \\
28: $pm2_1b$ (245) &  29: $pb2_1m$ (24)   & 26: $Pmc2_1$ \\
24: $pma2$ (0) &  31: $pm2a$ (10) &   28: $Pma2$ \\
40: $pmam$ (5) &  41: $pmma$ (154) &   51: $Pmma$ \\
\hline
\hline
\end{tabular}
\caption{List of all cases where two different layer group types are mapped to the same space group type upon AA stacking of the 2D material. The number of monolayers in the C2DB is indicated in parenthesis after the Hermann-Mauguin symbol of each layer group type with the tolerance parameter set to $10^{-1}$\AA. In Fig.~\ref{fig:comparison} some illustrative examples of crystal structures from the relevant layer group types are shown.}
\label{tab:spgmapping}
\end{table}

\section{Conclusions}
We have introduced an algorithm to determine the layer group of a crystal structure with a two-dimensional (2D) periodic lattice, and applied it to 15\,000+ atomically thin crystals stored in the Computational 2D Materials Database (C2DB). A symmetry analysis of the 2D materials in C2DB revealed a rather inhomogeneous distribution of layer groups with several layer groups only sparsely populated and 15 groups not represented at all. It would be interesting for future 2D materials discovery projects to focus on crystals from these layer groups. We have shown that classification of 2D materials by the space group of the AA-stacked bulk crystal is ambiguous for a significant fraction (18\%) of the C2DB materials, due to the inability of the space group to distinguish between the in-plane and out-of-plane directions. Specifically, there are nine pairs of layer group types, which cannot be distinguished by the space group of the AA-stacked bulk crystal. 

The method to determine the layer group is available as part of the open source \texttt{Spglib} library. The algorithm will be useful for researchers working with 2D crystal structures in general, and will facilitate the application of various theories and concepts for 2D materials that build on the layer group symmetry~\cite{PhysRevLett.115.126803,PhysRevLett.130.146801,delaFlor:ug5030,PhysRevLett.130.146801,PhysRevB.99.075105,PhysRevResearch.3.013052,PhysRevB.96.125127,Lazic2022}.

\begin{acknowledgments}
We acknowledge funding from the European Research Council (ERC) under the European Union’s Horizon 2020 research and innovation program Grant No. 773122 (LIMA) and Grant agreement No. 951786 (NOMAD CoE). This work was also supported by the Basic Science Center Project of NSFC (Grant No. 52388201). K. S. T. is a Villum Investigator supported by VILLUM FONDEN (grant no. 37789).

\end{acknowledgments}

\appendix

\section{Classification of layer group}

Below we summarize the mathematical structure and classification of the layer groups.
We mainly follow the terminologies in Refs.~\onlinecite{ITA} and \onlinecite{ITE}.
We classify the layer groups according to the point groups (Appendix.~\ref{sec:classification-pointgroup}), Bravais lattice (Appendix.~\ref{sec:classification-lattice}), and both (Appendix.~\ref{sec:classification-arithmetic}).

\subsection{\label{sec:classification-pointgroup}Point group of layer group}

Two point groups, $\mathcal{P}$ and $\mathcal{P}'$, belong to the same geometric crystal class if they are conjugate by an invertible matrix $\bm{P}$, i.e. $\bm{P}^{-1} \mathcal{P} \bm{P} = \mathcal{P}'$.
Two layer groups, $\mathrm{L}$ and $\mathrm{L}'$, with point groups $\mathcal{P}$ and $\mathcal{P}'$ belong to the same geometric crystal class if and only if $\mathcal{P}$ and $\mathcal{P}'$ belong to the same geometric crystal class.
Because the point groups of layer groups are three-dimensional, geometric crystal classes of layer groups are classified with the same notation as those of space groups.

Two point groups belong to the same crystal system if and only if the sets of Bravais type of lattices which these point groups leave invariant, coincide.
Crystal systems of layer groups are also classified with the same notation as those of space groups: triclinic, monoclinic, orthorhombic, tetragonal, trigonal, and hexagonal.
Note that a cubic crystal system does not exist for layer groups because the point groups of layer groups should preserve the 2D lattice plane (the point group of the cubic crystal system would contain an operation that rotates the structure out of the lattice plane).
Table~\ref{tab:geometric_layer} shows the classification of point groups for layer groups.

\begin{table}[tb]
  \centering
  \caption{
    Classification of the point groups of layer groups.
    Names and symbols of the crystal systems and geometric crystal classes are based on Table~3.2.1.4 of Ref.~\onlinecite{ITA}.
    The Laue class classifies the geometric crystal classes by ignoring inversion.
  }
  \label{tab:geometric_layer}
  \begin{ruledtabular}
  \begin{tabular}{l|ll}
  Crystal system & Laue class & Geometric crystal classes            \\ \hline
  Triclinic           & $\bar{1}$  & $1$, $\bar{1}$                     \\ \hline
  Monoclinic          & $2/m$           & $2$, $m$, $2/m$                         \\ \hline
  Orthorhombic        & $mmm$           & $222$, $2mm$, $mmm$                     \\ \hline
  Tetragonal          & $4/m$           & $4$, $\bar{4}$, $4/m$              \\
                      & $4/mmm$         & $422$, $\bar{4}2m$, $4mm$, $4/mmm$ \\ \hline
  Trigonal            & $\bar{3}$  & $3$, $\bar{3}$                     \\
                      & $\bar{3}m$ & $32$, $3m$, $\bar{3}m$             \\ \hline
  Hexagonal           & $6/m$           & $6$, $\bar{6}$, $6/m$              \\
                      & $6/mmm$         & $622$, $\bar{6}2m$, $6mm$, $6/mmm$ \\
  \end{tabular}
  \end{ruledtabular}
\end{table}

\subsection{\label{sec:classification-lattice}Translation lattice of layer group}

For a lattice $\mathbf{L}$, a Bravais group $\mathcal{B}(\mathbf{L})$ is a set of isometry mappings that preserve $\mathbf{L}$. Two lattices $\mathbf{L}$ and $\mathbf{L}'$ belong to the same Bravais type of lattices if and only if their Bravais groups are conjugate by an unimodular matrix $\bm{P}$, i.e. $\bm{P}^{-1} \mathcal{B}(\mathbf{L}) \bm{P} = \mathcal{B}(\mathbf{L}')$. Because translation lattices of layer groups are two-dimensional, Bravais types of lattices of layer groups are classified with the same notation as those of plane groups.

Two lattices $\mathbf{L}$ and $\mathbf{L}'$ belong to the same lattice system if and only if their Bravais groups belong to the same geometric crystal class. The corresponding point group for the geometric crystal class is called a holohedry. Lattice systems of layer groups are also classified with the same notation as those of plane groups: Oblique, Rectangular, Square, and Hexagonal. Table~\ref{tab:bravais_layer} shows the classification of translation lattices of layer groups.

\begin{table}[tb]
  \centering
  \caption{
    Classification of translation lattices of layer groups.
    Names and symbols of lattice systems are based on Table~3.1.1.1 of Ref.~\onlinecite{ITA}.
    Symbols of Bravais types of lattices are based on Table~3.1.2.1 of Ref.~\onlinecite{ITA}.
  }
  \label{tab:bravais_layer}
  \begin{ruledtabular}
  \begin{tabular}{l|l|l}
  Lattice system             & Holohedry & Bravais type  \\ \hline
  Oblique (monoclinic) $m$   & $2/m$     & $mp$          \\ \hline
  Rectangular                & $mmm$     & $op$          \\ 
  (orthorhombic) $o$         &           & $oc$          \\ \hline
  Square (tetragonal) $t$    & $4/mmm$   & $tp$          \\ \hline
  Hexagonal $h$              & $6/mmm$   & $hp$          \\
  \end{tabular}
  \end{ruledtabular}
\end{table}

\subsection{\label{sec:classification-arithmetic}Arithmetic-geometric crystal class}

Let $\mathcal{T}_{i}$ and $\mathcal{P}_{i}$ be a translation subgroup and a point group of layer groups $\mathrm{L}_{i} \, (i=1, 2)$.
Two layer groups $\mathrm{L}_{1}$ and $\mathrm{L}_{2}$ belong to the same arithmetic-geometric crystal class~\cite{Eick2005} if $\mathcal{P}_{1}$ and $\mathcal{P}_{2}$ are conjugate by a basis transformation from $\mathcal{T}_{1}$ to $\mathcal{T}_{2}$ and a transformation along the non-periodic axis.
Table~\ref{tab:arithmetic-geometric_layer} shows the classification of layer groups by the arithmetic-geometric crystal classes.

\begin{table}[tb]
  \centering
  \caption{
    Arithmetic-geometric crystal classes for layer groups.
    Each arithmetic-geometric crystal class is represented by a layer group in ITE whose sequential number is the smallest among the belonging arithmetic-geometric crystal class.
  }
  \label{tab:arithmetic-geometric_layer}
  \begin{ruledtabular}
  \begin{tabular}{l|l|l}
  Crystal class/lattice system & Geometric     & Arithmetic-   \\
                               & crystal class & geometric     \\
                               &               & crystal class \\ \hline
  Triclinic/oblique            & $1$                     & $p1$ (1)                           \\
                              & $\bar{1}$          & $p\bar{1}$ (2)                \\ \hline
  Monoclinic/oblique           & $2$                     & $p112$ (3)                         \\
                              & $m$                     & $p11m$ (4)                         \\
                              & $2/m$                   & $p112/m$ (6)                       \\ \hline
  Monoclinic/rectangular       & $2$                     & $p211$ (8)                         \\
                              &                         & $c211$ (10)                        \\
                              & $m$                     & $pm11$ (11)                        \\
                              &                         & $cm11$ (13)                        \\
                              & $2/m$                   & $p2/m11$ (14)                      \\
                              &                         & $c2/m11$ (18)                      \\ \hline
  Orthorhombic/rectangular     & $222$                   & $p222$ (19)                        \\
                              &                         & $c222$ (22)                        \\
                              & $mm2$                   & $pmm2$ (23)                        \\
                              &                         & $cmm2$ (26)                        \\
                              & ($m2m$)                 & $pm2m$ (27)                        \\
                              &                         & $cm2m$ (35)                        \\
                              & $mmm$                   & $pmmm$ (37)                        \\
                              &                         & $cmmm$ (47)                        \\ \hline
  Tetragonal/square            & $4$                     & $p4$ (49)                         \\ 
                              & $\bar{4}$          & $p\bar{4}$ (50)               \\ 
                              & $4/m$                   & $p4/m$ (51)                        \\ 
                              & $422$                   & $p422$ (53)                        \\ 
                              & $4mm$                   & $p4mm$ (55)                        \\ 
                              & $\bar{4}2m$        & $p\bar{4}2m$ (57)             \\
                              & ($\bar{4}m2$)      & $p\bar{4}m2$ (59)             \\ 
                              & $4/mmm$                 & $p4/mmm$ (61)                      \\ \hline
  Trigonal/hexagonal           & $3$                     & $p3$ (65)                         \\
                              & $\bar{3}$          & $p\bar{3}$ (66)               \\
                              & $312$                   & $p312$ (67)                        \\
                              & ($321$)                 & $p321$ (68)                        \\
                              & $3m1$                   & $p3m1$ (69)                        \\
                              & ($31m$)                 & $p31m$ (70)                        \\
                              & $\bar{3}1m$        & $p\bar{3}1m$ (71)             \\
                              & ($\bar{3}m1$)      & $p\bar{3}m1$ (72)             \\ \hline
  Hexagonal/hexagonal          & $6$                     & $p6$ (73)                         \\
                              & $\bar{6}$          & $p\bar{6}$ (74)               \\
                              & $6/m$                   & $p6/m$ (75)                        \\
                              & $622$                   & $p622$ (76)                        \\
                              & $6mm$                   & $p6mm$ (77)                        \\
                              & $\bar{6}m2$        & $p\bar{6}m2$ (78)             \\
                              & ($\bar{6}2m$)      & $p\bar{6}2m$ (79)             \\
                              & $6/mmm$                 & $p6/mmm$ (80)                      \\
  \end{tabular}
  \end{ruledtabular}
\end{table}

\section{DELAUNAY REDUCTION}
\label{sec:delaunay}
\begin{figure}
    \centering
    \resizebox{\columnwidth}{!}{\begin{tikzpicture}[font=\sffamily\large, line width = 1pt]
    \definecolor{fill_B}{RGB}{255,242,174}
    \definecolor{fill_C}{RGB}{230,245,201}
    \definecolor{draw_C}{RGB}{60,143,62}
    \foreach \T/\X/\C in {A/-6/fill_C, B/-3/fill_B, C/0/fill_C, D/3/fill_B} {
        \fill[\C] (\X, -5) rectangle +(3,10);
        \fill[white] (\X + 1.5, 0) circle (1.5cm);
        \draw [gray, very thin] (\X + 1, 0.8) circle (2pt);
        \draw [gray, very thin] (\X + 2, -0.8) circle (2pt);
        \draw (\X + 1.5, -3) node {\T};
    }
    
    \draw[draw_C, line width = 2pt] (0,-5) -- +(0,10);
    \draw[draw_C, line width = 2pt] (3,-5) -- +(0,10);
    \draw[draw_C, line width = 2pt] (1.5,0) circle (1.5cm);

    \fill[white] (-6,4.65) rectangle +(12,0.1);
    \fill[white] (-6,4.85) rectangle +(12,0.1);
    \fill[white] (-6,-4.65) rectangle +(12,-0.1);
    \fill[white] (-6,-4.85) rectangle +(12,-0.1);

    \draw[-stealth] (0,0) -- (-3,0) node[anchor=east]  {$\textbf{b}_\textsf{\scriptsize 1}$};
    \draw[-stealth,dashed] (0,0) -- (3,0) node[anchor=west] {-$\textbf{b}_\textsf{\scriptsize 1}$};
    \draw[-stealth] (0,0) -- (1,0.8) node[anchor=west]  {$\textbf{b}_\textsf{\scriptsize 2}$};
    \draw[-stealth] (0,0) -- (2,-0.8) node[anchor=west]  {$\textbf{b}_\textsf{\scriptsize 3}$};
    \draw[-stealth,dotted] (0,0) -- (-1,1.6) node[anchor=south]  {$\textbf{b}_\textsf{\scriptsize 1}\textsf{+2}\textbf{b}_\textsf{\scriptsize 2}$};
    \draw[-stealth,dotted] (0,0) -- (-1,-0.8) node[anchor=east]  {-$\textbf{b}_\textsf{\scriptsize 2}$};
    
    \fill[black] (0,0) circle (2pt);
    \draw (-0.05,-0.05) node[anchor=north west] {$\textit{O}$};

    \fill[black] (6.3,0) circle (1pt);
    \fill[black] (6.5,0) circle (1pt);
    \fill[black] (6.7,0) circle (1pt);
    \fill[black] (-6.3,0) circle (1pt);
    \fill[black] (-6.5,0) circle (1pt);
    \fill[black] (-6.7,0) circle (1pt);
\end{tikzpicture}}
    \caption{Schematic diagram of 2D Delaunay reduction. The solid arrows denote the vectors at an intermediate step during the Delaunay reduction. Since the angle between $\mathbf{b}_2$ and $\mathbf{b}_3$ is smaller than $90^\circ$, the next step will transform $\mathbf{b}_1$ and $\mathbf{b}_2$ to $\mathbf{b}_1 + 2\mathbf{b}_2$ and $-\mathbf{b}_2$, respectively. The transformed vectors are denoted by dotted arrows. If the 2D lattice is spanned by $\mathbf{b}_1$ and a vector perpendicular to the paper, such transformation breaks the lattice plane and is not allowed. The Delaunay reduction will stop at the previous step. The above-mentioned case appears when $\mathbf{b}_2$ lies in the white circles. The Delaunay reduction can be fully applied when $\mathbf{b}_2$ lies in the rest areas (edge included). Green and yellow indicate $\mathbf{b}_1$ finally being transformed to $\mathbf{b}_1$ and $-\mathbf{b}_1$, respectively.}
    \label{fig:2d-delaunay}
\end{figure}
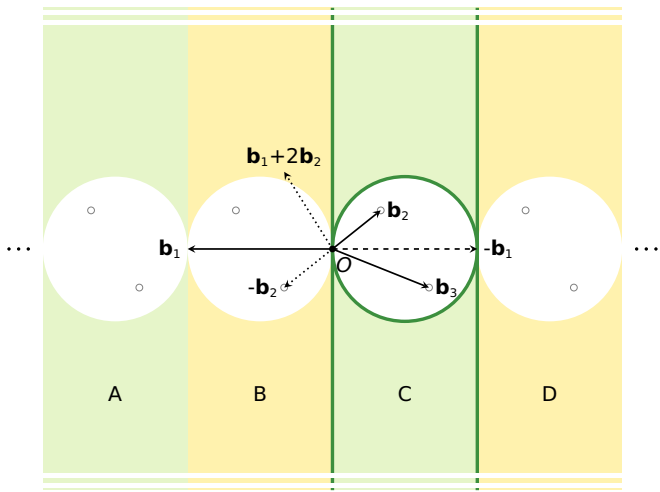

Given an $n$-dimensional ($n$D) lattice, the goal of Delaunay reduction is to find the shortest basis vectors $\mathbf{b}_i(1 \leq i \leq n)$ spanning the lattice. In practice, the algorithm achieves that by minimizing \begin{equation}
\label{eq:target-to-minimize}
    \sum_{i=1}^{n+1} \mathbf{b}_i^2,
\end{equation}
where
\begin{equation}
    \mathbf{b}_{n+1} = - \sum_{i=1}^n \mathbf{b}_i.
\end{equation}

For $n=2$, we start by selecting two basis vectors $\mathbf{b}_1$ and $\mathbf{b}_2$. The extended basis $\mathbf{b}_3=-\mathbf{b}_1-\mathbf{b}_2$. Check the 3 scalar products $\mathbf{b}_i \cdot \mathbf{b}_j (i \ne j)$ one after the other. Any time $\mathbf{b}_i \cdot \mathbf{b}_j > 0$, the transformation
\begin{equation}
\begin{aligned}
    &\mathbf{b}'_i& &=& - &\mathbf{b}_i \\
    &\mathbf{b}'_j& &=&   &\mathbf{b}_j \\
    &\mathbf{b}'_k& &=&   &\mathbf{b}_k + 2 \mathbf{b}_i
\end{aligned}
\end{equation}
is performed, where $k \ne i,j$. After the transformation, $\sum {\mathbf{b}'_{i'}}^2 = \sum \mathbf{b}_{i'}^2 - 4 \mathbf{b}_i \cdot \mathbf{b}_j < \sum \mathbf{b}_{i'}^2$. Then we check $\mathbf{b}'_1 \cdot \mathbf{b}'_2$, $\mathbf{b}'_1 \cdot \mathbf{b}'_3$ and $\mathbf{b}'_2 \cdot \mathbf{b}'_3$, and so on. The loop stops when all $\mathbf{b}_i \cdot \mathbf{b}_j \leq 0$, i.e., the angles between $\mathbf{b}_1,\mathbf{b}_2,\mathbf{b}_3$ all become non-acute. 2D Delaunay reduction is applied to the basis vectors forming the oblique face of a monoclinic cell and works well on bulk systems. For the monoclinic/oblique cell, the target vectors to be reduced are the lattice vectors $\mathbf{a}$ and $\mathbf{b}$, and the procedure has no difference from the bulk case. However, the oblique face of a monoclinic/rectangular cell is composed of $\mathbf{b}_1=\mathbf{b}$ and $\mathbf{b}_2=\mathbf{c}$. As shown in Fig.~\ref{fig:2d-delaunay}, the basis vectors of the periodic plane are $\mathbf{b}_1$ and the unique axis $\mathbf{a}$ which is vertical to the paper. Allowed transformations for $\mathbf{b}_1$ are $\mathbf{b}_1 \to \mathbf{b}_1$ and $\mathbf{b}_1 \to -\mathbf{b}_1$. When the angle between $\mathbf{b}_2$ and $\mathbf{b}_3$ is acute, Delaunay reduction requires $\mathbf{b}_1 \to \mathbf{b}_1 + 2 \mathbf{b}_2$ and $\mathbf{b}_2 \to -\mathbf{b}_2$. The new vectors break the lattice plane, which is not permitted (dotted arrows in Fig.~\ref{fig:2d-delaunay}). For certain input cell, candidates for $\mathbf{b}_2, \mathbf{b}_3$ are predetermined (represented by hollow points in Fig.~\ref{fig:2d-delaunay}). When the points lay on the colored areas (including edges), $\forall \mathbf{b}_i \cdot \mathbf{b}_j \leq 0$ fulfills after Delaunay reduction. $\mathbf{b}_2$ with the initial position at the green areas will be sent to the green area of region C. $\mathbf{b}_2$ in the yellow areas will finally deposit in region B, with $\mathbf{b}_1$ being switched to $-\mathbf{b}_1$. If the input $\mathbf{b}_2$ lies in any of the white circles, at least one angle between the vectors will be acute. In the algorithm, it is set to be the one between $\mathbf{b}_2$ and $\mathbf{b}_3$. After reduction, the non-lattice vector $\mathbf{c}$ and the shortest two vectors among $\mathbf{b}_1$, $\mathbf{b}_2$ and $\mathbf{b}_3$ are selected to build the bulk or monoclinic/oblique cell. While for a monoclinic/rectangular cell, one of $\pm \mathbf{b}_1$ must be chosen to maintain the periodic plane. The other two vectors are $\mathbf{a}$ and the shortest in $\mathbf{b}_2$ and $\mathbf{b}_3$.

The workflow for $n=3$ is similar. The transformation for $\mathbf{b}_i \cdot \mathbf{b}_j > 0 \;(i,j=1,\ldots,4; \;i \ne j)$ reads
\begin{equation}
\begin{aligned}
    &\mathbf{b}'_i& &=& - &\mathbf{b}_i \\
    &\mathbf{b}'_j& &=&   &\mathbf{b}_j \\
    &\mathbf{b}'_k& &=&   &\mathbf{b}_k + \mathbf{b}_i \\
    &\mathbf{b}'_l& &=&   &\mathbf{b}_l + \mathbf{b}_i
\end{aligned}
\end{equation}

For layer systems, when the non-lattice vector $\mathbf{c}$ lies in the intersection of an infinitely long prism and a sphere, the Delaunay reduction cannot be fully applied. The angle between $\mathbf{b}_3$ and $\mathbf{b}_4$ will be less than $90^\circ$. The two basis vectors that form the cross-section of the prism are
\begin{equation}
\begin{aligned}
    &\mathbf{p}_1 = \frac{(\mathbf{b}_1 + \mathbf{b}_2) \cdot \mathbf{b}_2}
        {(\mathbf{b}_2 \times \mathbf{b}_1) \cdot \mathbf{e}_z}
    (\mathbf{e}_z \times \mathbf{b}_1) \text{ and }, \\
    &\mathbf{p}_2 = \frac{(\mathbf{b}_1 + \mathbf{b}_2) \cdot \mathbf{b}_1}
        {(\mathbf{b}_2 \times \mathbf{b}_1) \cdot \mathbf{e}_z}
    (\mathbf{e}_z \times \mathbf{b}_2),
\end{aligned}
\end{equation}
where $\mathbf{e}_z = (0, 0, 1)$, $\mathbf{p}_1 \bot \mathbf{b}_1$, $\mathbf{p}_2 \bot \mathbf{b}_2$, and $\mathbf{p}_1 + \mathbf{p}_2 = - \mathbf{b}_1 - \mathbf{b}_2$. The prism is infinite along the $\pm z$ direction. The origin of the sphere is $-\frac{1}{2} (\mathbf{b}_1 + \mathbf{b}_2)$ and the radius is $-\frac{1}{2} \lvert \mathbf{b}_1 + \mathbf{b}_2 \rvert$.

\section{CONVENTIONAL CELLS}
\label{sec:conv-cell}
The standard crystallographic coordinate system (CCS) choices defined in Refs.~\onlinecite{ITA} and \onlinecite{ITE} impose constraints on the choices and the orientations of the basis vectors, and are firstly consulted for building the conventional cells. Additional metric rules are added when the standard choices cannot uniquely determine the cell. The criteria used by \texttt{Spglib} are summarized in Fig.~\ref{fig:workflow-conv-cell}.


For triclinic lattices, a Niggli cell~\cite{doi:10.1080/11035892909447060} is chosen to be the conventional cell by virtue of its uniqueness. For bulk crystals (3D lattice), the Niggli conditions include $\lvert \mathbf{a}_\mathrm{c} \rvert \leq
\lvert \mathbf{b}_\mathrm{c} \rvert \leq \lvert \mathbf{c}_\mathrm{c} \rvert$ and $\lvert \mathbf{a}_\mathrm{c} \cdot \mathbf{c}_\mathrm{c} \rvert \leq \lvert \mathbf{a}_\mathrm{c} \cdot \mathbf{b}_\mathrm{c} \rvert$ if $\lvert \mathbf{b}_\mathrm{c} \rvert = \lvert \mathbf{c}_\mathrm{c} \rvert$ for bulk systems. The Niggli reduction will swap $\mathbf{b}_\mathrm{c}$ and $\mathbf{c}_\mathrm{c}$ when the aforementioned conditions are not met. For layer systems with the non-lattice vector $\mathbf{c}_\mathrm{c}$, such action is not allowed. In this case, the resulting cell will thus differ from the conventional Niggli cell by a rotation.

The monoclinic crystal system is subdivided into monoclinic/oblique and monoclinic/rectangular according to the 2D Bravais lattice. The unique two-fold axis is vertical to the plane and is set to be parallel to $\mathbf{c}_\mathrm{c}$ for the former, while for the latter the axis is in-plane and chosen to be along $\mathbf{a}_\mathrm{c}$. 2D Delaunay reduction is conducted for the oblique face of the cell. As shown in Appendix.~\ref{sec:delaunay}, this procedure cannot be completely performed for the monoclinic/rectangular system. In this case, both the shape and the orientation of the resulting cell will differ from the conventional cell that would have been obtained for a monoclinic bulk crystal.

The conventional cells of the orthorhombic groups are cuboids. The standard CCS choice requires the axes of certain symmetry operations to be perpendicular to the faces of the cuboids. When multiple faces share the same symmetry diagram up to an origin shift, permutation of the cell vectors will not influence the CCS choice. To find out which basis vectors can be swapped, the Euclidean normalizer of the space/layer group is utilized. The Euclidean normalizer of a group $\mathcal{G}$ is the group $\mathcal{N_E}$ consisting of all Euclidean transformations which map $\mathcal{G}$ onto itself by conjugation
\begin{equation}
    \mathcal{N_E(G)} \equiv \left\{ s \in \mathcal{E} \mid s^{-1}\mathcal{G}s = \mathcal{G} \right\},
\end{equation}
where $\mathcal{E}$ is the group of all Euclidean transformations~\cite{ITA}. $\mathcal{G}$ is a normal subgroup of $\mathcal{N_E}$, and the factor group $\mathcal{N_E(G) / G}$ classifies the transformations. Each element $s\mathcal{G}$ of the factor group represents a different kind of CCS transformation (axis permutation, origin shift, etc.) that does not alter the CCS choice. We are interested in all the $s\mathcal{G}$ which contains one of the six right-handed axis permutations $\mathbf{a}\mathbf{b}\mathbf{c}$, $\mathbf{b}\mathbf{a}\bar{\mathbf{c}}$, $\mathbf{c}\mathbf{a}\mathbf{b}$, $\bar{\mathbf{c}}\mathbf{b}\mathbf{a}$, $\mathbf{b}\mathbf{c}\mathbf{a}$ and $\mathbf{a}\bar{\mathbf{c}}\mathbf{b}$. They can be obtained by exploiting the group-subgroup relationship revealed in section 3.5.2.1 of Ref.~\onlinecite{ITA}
\begin{equation}
    \mathcal{G} \le \mathcal{K(G)} \le \mathcal{L(G)} \le \mathcal{N_E(G)},
\end{equation}
where $\mathcal{K(G)}$ is an intermediate group that retains the linear part of $\mathcal{G}$ and the translation part of $\mathcal{N_E(G)}$. $\mathcal{L(G)} = \mathcal{K(G)} \otimes \left\{ 1, \bar{1} \right\}$ if $\mathcal{G}$ does not have inversion symmetry while $\mathcal{N_E(G)}$ is centrosymmetric, else $\mathcal{L(G)} = \mathcal{K(G)}$. The group $\mathcal{N_E(G) / L(G)}$ factors out choices related to origin shift and inversion which flips the handedness. The index $n_l = \lvert \mathcal{N_E(G) / L(G)} \rvert$ indicates the number of axes permutation types that preserves the CCS choice. For space groups, when $n_l = 6$, the six different axis permutations do not influence the CCS choice. The order of the three basis vectors can be arbitrarily chosen. \texttt{Spglib} requires $\lvert \mathbf{a}_\mathrm{c} \rvert \leq
\lvert \mathbf{b}_\mathrm{c} \rvert \leq \lvert \mathbf{c}_\mathrm{c} \rvert$. Space groups $P222$, $F222$, $Ibca$, etc. belong to this case. Space group $Pbca$ is the only one with $n_l = 3$, where $\mathbf{a}\mathbf{b}\mathbf{c}$, $\mathbf{c}\mathbf{a}\mathbf{b}$ and $\mathbf{b}\mathbf{c}\mathbf{a}$ preserves the standard CCS choice, while $\mathbf{b}\mathbf{a}\bar{\mathbf{c}}$, $\bar{\mathbf{c}}\mathbf{b}\mathbf{a}$ and $\mathbf{a}\bar{\mathbf{c}}\mathbf{b}$ changes the choice to another. \texttt{Spglib} chooses $\mathbf{a}_\mathrm{c}$ to be the shortest vector, the orientation of $\mathbf{b}_\mathrm{c}$ and $\mathbf{c}_\mathrm{c}$ will be settled accordingly. $n_l = 2$ means the standard choice fixes certain symmetry axis being parallel to $\mathbf{c}_\mathrm{c}$. $\mathbf{a}_\mathrm{c}$ and $\mathbf{b}_\mathrm{c}$ cannot be determined by symmetry. \texttt{Spglib} adds the constraint $\lvert \mathbf{a}_\mathrm{c} \rvert \leq
\lvert \mathbf{b}_\mathrm{c} \rvert$. Typical space groups are $P222_1$, $C222$, etc. The rest orthorhombic space groups like $Pmc2_1$ have $n_l = 1$, of which the standard choices uniquely determine the basis of the conventional cells. For layer groups, $n_l \le 2$ due to $\mathbf{c}_\mathrm{c}$ being the non-lattice vector. 

For bulk, Table 3.5.2.4 of Ref.~\onlinecite{ITA} lists $n_l$ for centrosymmetric $\mathcal{N_E(G)}$, and $n_k = \lvert \mathcal{N_E(G) / K(G)} \rvert$ for noncentrosymmetric $\mathcal{N_E(G)}$. For simplicity, we use $n_l$ to denote the two symbols. Different metric parameter conditions may lead to different $\mathcal{N_E(G)}$, and the one with the highest symmetry should be consulted. For layer, $n_l$ can be deduced from the table by consulting the line with the highest symmetry under the metric conditions $a \ne c$ and $b \ne c$. It can also be found in Ref.~\onlinecite{VanLeeuwen:dm5060}.

For space groups, additional correction matrices must be applied for monoclinic and orthorhombic conventional cells with `A', `B' or body-centering types, as the standard choices only possess `C' centering lattices. The correction matrices are listed in Ref.~\onlinecite{spglibv1}. This step is not necessary for layer groups. 

\section{HALL SYMBOLS}
\label{sec:hall-symbols}
Hall symbols are used to represent the different CCS choices of space and layer groups in \texttt{Spglib}. Compared to Hermann–Mauguin symbols, Hall symbols take the origin of each symmetry operation into consideration. A unique symmetry group can be deduced from a Hall symbol without additional information. The 530 Hall symbols for different choices of space groups are tabulated in table A1.4.2.7 of Ref.~\onlinecite{ITB}. As there does not exist a standard table for layer group Hall symbols, we have produced Table~\ref{tab:hall-symbol-table} as a reference.

The first column of the table gives the layer group number. If there are multiple CCS choices for one layer group type, the group number is followed by the axis codes of the choice. The first row of each group number always corresponds to the standard choice defined in chapter 4 of Ref.~\onlinecite{ITE}. The non-standard choices are generated according to table 1.2.6.1 and chapter 4 of Ref.~\onlinecite{ITE}. The second to the fourth column lists the Hermann–Mauguin entries, Hall entries suitable for computer processing and the Hall symbols, respectively. The detailed meaning of the Hall symbols is described in Ref.~\onlinecite{ITB}, except we use symbols with the first letter upper and lowercase to distinguish the space and layer groups, respectively.

\begin{table*}
    \centering
    \renewcommand*{\arraystretch}{1.2}
    \resizebox*{!}{0.92\textheight}{
        \begin{tabular}{|p{0.12\textwidth}|p{0.12\textwidth}|p{0.12\textwidth}|p{0.12\textwidth}|l|p{0.12\textwidth}|p{0.12\textwidth}|p{0.12\textwidth}|p{0.12\textwidth}|}
\cline{1-4} \cline{6-9}
n:c     & H-M entry  & Hall entry   & Hall symbol                              & &n:c     & H-M entry  & Hall entry   & Hall symbol                             \\
\cline{1-4} \cline{6-9}
1       & p 1        & p 1          & $\mathrm{p\ 1}$                          & & 34:b-ac & p 2 a n    & p -2ab 2     & $\mathrm{p\ \bar{2}_{ab}\ 2}$           \\
2       & p -1       & -p 1         & $\mathrm{\bar{p}\ 1}$                    & & 35      & c m 2 m    & c -2 -2      & $\mathrm{c\ \bar{2}\ \bar{2}}$          \\
3:c     & p 1 1 2    & p 2          & $\mathrm{p\ 2}$                          & & 35:b-ac & c 2 m m    & c -2 2       & $\mathrm{c\ \bar{2}\ 2}$                \\
4:c     & p 1 1 m    & p -2         & $\mathrm{p\ \bar{2}}$                    & & 36      & c m 2 e    & c -2a -2a    & $\mathrm{c\ \bar{2}_{a}\ \bar{2}_{a}}$  \\
5:c1    & p 1 1 a    & p -2a        & $\mathrm{p\ \bar{2}_{a}}$                & & 36:b-ac & c 2 m e    & c -2a 2      & $\mathrm{c\ \bar{2}_{a}\ 2}$            \\
5:c2    & p 1 1 n    & p -2ab       & $\mathrm{p\ \bar{2}_{ab}}$               & & 37      & p m m m    & -p 2 2       & $\mathrm{\bar{p}\ 2\ 2}$                \\
5:c3    & p 1 1 b    & p -2b        & $\mathrm{p\ \bar{2}_{b}}$                & & 38      & p m a a    & -p 2a 2      & $\mathrm{\bar{p}\ 2_{a}\ 2}$            \\
6:c     & p 1 1 2/m  & -p 2         & $\mathrm{\bar{p}\ 2}$                    & & 38:b-ac & p b m b    & -p 2b 2b     & $\mathrm{\bar{p}\ 2_{b}\ 2_{b}}$        \\
7:c1    & p 1 1 2/a  & -p 2a        & $\mathrm{\bar{p}\ 2_{a}}$                & & 39      & p b a n    & -p 2ab 2b    & $\mathrm{\bar{p}\ 2_{ab}\ 2_{b}}$       \\
7:c2    & p 1 1 2/n  & -p 2ab       & $\mathrm{\bar{p}\ 2_{ab}}$               & & 40      & p m a m    & -p 2 2a      & $\mathrm{\bar{p}\ 2\ 2_{a}}$            \\
7:c3    & p 1 1 2/b  & -p 2b        & $\mathrm{\bar{p}\ 2_{b}}$                & & 40:b-ac & p b m m    & -p 2 2b      & $\mathrm{\bar{p}\ 2\ 2_{b}}$            \\
8:a     & p 2 1 1    & p 2x         & $\mathrm{p\ 2^{x}}$                      & & 41      & p m m a    & -p 2a 2a     & $\mathrm{\bar{p}\ 2_{a}\ 2_{a}}$        \\
8:b     & p 1 2 1    & p 2y         & $\mathrm{p\ 2^{y}}$                      & & 41:b-ac & p m m b    & -p 2b 2      & $\mathrm{\bar{p}\ 2_{b}\ 2}$            \\
9:a     & p 21 1 1   & p 2xa        & $\mathrm{p\ 2^{x}_{a}}$                  & & 42      & p m a n    & -p 2ab 2     & $\mathrm{\bar{p}\ 2_{ab}\ 2}$           \\
9:b     & p 1 21 1   & p 2yb        & $\mathrm{p\ 2^{y}_{b}}$                  & & 42:b-ac & p b m n    & -p 2ab 2ab   & $\mathrm{\bar{p}\ 2_{ab}\ 2_{ab}}$      \\
10:a    & c 2 1 1    & c 2x         & $\mathrm{c\ 2^{x}}$                      & & 43      & p b a a    & -p 2a 2b     & $\mathrm{\bar{p}\ 2_{a}\ 2_{b}}$        \\
10:b    & c 1 2 1    & c 2y         & $\mathrm{c\ 2^{y}}$                      & & 43:b-ac & p b a b    & -p 2b 2ab    & $\mathrm{\bar{p}\ 2_{b}\ 2_{ab}}$       \\
11:a    & p m 1 1    & p -2x        & $\mathrm{p\ \bar{2}^{x}}$                & & 44      & p b a m    & -p 2 2ab     & $\mathrm{\bar{p}\ 2\ 2_{ab}}$           \\
11:b    & p 1 m 1    & p -2y        & $\mathrm{p\ \bar{2}^{y}}$                & & 45      & p b m a    & -p 2a 2ab    & $\mathrm{\bar{p}\ 2_{a}\ 2_{ab}}$       \\
12:a    & p b 1 1    & p -2xb       & $\mathrm{p\ \bar{2}^{x}_{b}}$            & & 45:b-ac & p m a b    & -p 2b 2a     & $\mathrm{\bar{p}\ 2_{b}\ 2_{a}}$        \\
12:b    & p 1 a 1    & p -2ya       & $\mathrm{p\ \bar{2}^{y}_{a}}$            & & 46      & p m m n    & -p 2ab 2a    & $\mathrm{\bar{p}\ 2_{ab}\ 2_{a}}$       \\
13:a    & c m 1 1    & c -2x        & $\mathrm{c\ \bar{2}^{x}}$                & & 47      & c m m m    & -c 2 2       & $\mathrm{\bar{c}\ 2\ 2}$                \\
13:b    & c 1 m 1    & c -2y        & $\mathrm{c\ \bar{2}^{y}}$                & & 48      & c m m e    & -c 2a 2      & $\mathrm{\bar{c}\ 2_{a}\ 2}$            \\
14:a    & p 2/m 1 1  & -p 2x        & $\mathrm{\bar{p}\ 2^{x}}$                & & 49      & p 4        & p 4          & $\mathrm{p\ 4}$                         \\
14:b    & p 1 2/m 1  & -p 2y        & $\mathrm{\bar{p}\ 2^{y}}$                & & 50      & p -4       & p -4         & $\mathrm{p\ \bar{4}}$                   \\
15:a    & p 21/m 1 1 & -p 2xa       & $\mathrm{\bar{p}\ 2^{x}_{a}}$            & & 51      & p 4/m      & -p 4         & $\mathrm{\bar{p}\ 4}$                   \\
15:b    & p 1 21/m 1 & -p 2yb       & $\mathrm{\bar{p}\ 2^{y}_{b}}$            & & 52:1    & p 4/n:1    & p 4 -1ab     & $\mathrm{p\ 4\ \bar{1}_{ab}}$           \\
16:a    & p 2/b 1 1  & -p 2xb       & $\mathrm{\bar{p}\ 2^{x}_{b}}$            & & 52:2    & p 4/n:2    & -p 4a        & $\mathrm{\bar{p}\ 4_{a}}$               \\
16:b    & p 1 2/a 1  & -p 2ya       & $\mathrm{\bar{p}\ 2^{y}_{a}}$            & & 53      & p 4 2 2    & p 4 2        & $\mathrm{p\ 4\ 2}$                      \\
17:a    & p 21/b 1 1 & -p 2xab      & $\mathrm{\bar{p}\ 2^{x}_{ab}}$           & & 54      & p 4 21 2   & p 4 2ab      & $\mathrm{p\ 4\ 2_{ab}}$                 \\
17:b    & p 1 21/a 1 & -p 2yab      & $\mathrm{\bar{p}\ 2^{y}_{ab}}$           & & 55      & p 4 m m    & p 4 -2       & $\mathrm{p\ 4\ \bar{2}}$                \\
18:a    & c 2/m 1 1  & -c 2x        & $\mathrm{\bar{c}\ 2^{x}}$                & & 56      & p 4 b m    & p 4 -2ab     & $\mathrm{p\ 4\ \bar{2}_{ab}}$           \\
18:b    & c 1 2/m 1  & -c 2y        & $\mathrm{\bar{c}\ 2^{y}}$                & & 57      & p -4 2 m   & p -4 2       & $\mathrm{p\ \bar{4}\ 2}$                \\
19      & p 2 2 2    & p 2 2        & $\mathrm{p\ 2\ 2}$                       & & 58      & p -4 21 m  & p -4 2ab     & $\mathrm{p\ \bar{4}\ 2_{ab}}$           \\
20      & p 21 2 2   & p 2 2a       & $\mathrm{p\ 2\ 2_{a}}$                   & & 59      & p -4 m 2   & p -4 -2      & $\mathrm{p\ \bar{4}\ \bar{2}}$          \\
20:b-ac & p 2 21 2   & p 2 2b       & $\mathrm{p\ 2\ 2_{b}}$                   & & 60      & p -4 b 2   & p -4 -2ab    & $\mathrm{p\ \bar{4}\ \bar{2}_{ab}}$     \\
21      & p 21 21 2  & p 2 2ab      & $\mathrm{p\ 2\ 2_{ab}}$                  & & 61      & p 4/m m m  & -p 4 2       & $\mathrm{\bar{p}\ 4\ 2}$                \\
22      & c 2 2 2    & c 2 2        & $\mathrm{c\ 2\ 2}$                       & & 62:1    & p 4/n b m:1 & p 4 2 -1ab   & $\mathrm{p\ 4\ 2\ \bar{1}_{ab}}$        \\
23      & p m m 2    & p 2 -2       & $\mathrm{p\ 2\ \bar{2}}$                 & & 62:2    & p 4/n b m:2 & -p 4a 2b     & $\mathrm{\bar{p}\ 4_{a}\ 2_{b}}$        \\
24      & p m a 2    & p 2 -2a      & $\mathrm{p\ 2\ \bar{2}_{a}}$             & & 63      & p 4/m b m  & -p 4 2ab     & $\mathrm{\bar{p}\ 4\ 2_{ab}}$           \\
24:b-ac & p b m 2    & p 2 -2b      & $\mathrm{p\ 2\ \bar{2}_{b}}$             & & 64:1    & p 4/n m m:1 & p 4 2ab -1ab & $\mathrm{p\ 4\ 2_{ab}\ \bar{1}_{ab}}$   \\
25      & p b a 2    & p 2 -2ab     & $\mathrm{p\ 2\ \bar{2}_{ab}}$            & & 64:2    & p 4/n m m:2 & -p 4a 2a     & $\mathrm{\bar{p}\ 4_{a}\ 2_{a}}$        \\
26      & c m m 2    & c 2 -2       & $\mathrm{c\ 2\ \bar{2}}$                 & & 65      & p 3        & p 3          & $\mathrm{p\ 3}$                         \\
27      & p m 2 m    & p -2 -2      & $\mathrm{p\ \bar{2}\ \bar{2}}$           & & 66      & p -3       & -p 3         & $\mathrm{\bar{p}\ 3}$                   \\
27:b-ac & p 2 m m    & p -2 2       & $\mathrm{p\ \bar{2}\ 2}$                 & & 67      & p 3 1 2    & p 3 2        & $\mathrm{p\ 3\ 2}$                      \\
28      & p m 21 b   & p -2b -2     & $\mathrm{p\ \bar{2}_{b}\ \bar{2}}$       & & 68      & p 3 2 1    & p 3 2"       & $\mathrm{p\ 3\ 2}$"                     \\
28:b-ac & p 21 m a   & p -2a 2a     & $\mathrm{p\ \bar{2}_{a}\ 2_{a}}$         & & 69      & p 3 m 1    & p 3 -2"      & $\mathrm{p\ 3\ \bar{2}}$"               \\
29      & p b 21 m   & p -2 -2b     & $\mathrm{p\ \bar{2}\ \bar{2}_{b}}$       & & 70      & p 3 1 m    & p 3 -2       & $\mathrm{p\ 3\ \bar{2}}$                \\
29:b-ac & p 21 a m   & p -2 2a      & $\mathrm{p\ \bar{2}\ 2_{a}}$             & & 71      & p -3 1 m   & -p 3 2       & $\mathrm{\bar{p}\ 3\ 2}$                \\
30      & p b 2 b    & p -2b -2b    & $\mathrm{p\ \bar{2}_{b}\ \bar{2}_{b}}$   & & 72      & p -3 m 1   & -p 3 2"      & $\mathrm{\bar{p}\ 3\ 2}$"               \\
30:b-ac & p 2 a a    & p -2a 2      & $\mathrm{p\ \bar{2}_{a}\ 2}$             & & 73      & p 6        & p 6          & $\mathrm{p\ 6}$                         \\
31      & p m 2 a    & p -2a -2a    & $\mathrm{p\ \bar{2}_{a}\ \bar{2}_{a}}$   & & 74      & p -6       & p -6         & $\mathrm{p\ \bar{6}}$                   \\
31:b-ac & p 2 m b    & p -2b 2      & $\mathrm{p\ \bar{2}_{b}\ 2}$             & & 75      & p 6/m      & -p 6         & $\mathrm{\bar{p}\ 6}$                   \\
32      & p m 21 n   & p -2ab -2    & $\mathrm{p\ \bar{2}_{ab}\ \bar{2}}$      & & 76      & p 6 2 2    & p 6 2        & $\mathrm{p\ 6\ 2}$                      \\
32:b-ac & p 21 m n   & p -2ab 2ab   & $\mathrm{p\ \bar{2}_{ab}\ 2_{ab}}$       & & 77      & p 6 m m    & p 6 -2       & $\mathrm{p\ 6\ \bar{2}}$                \\
33      & p b 21 a   & p -2a -2ab   & $\mathrm{p\ \bar{2}_{a}\ \bar{2}_{ab}}$  & & 78      & p -6 m 2   & p -6 2       & $\mathrm{p\ \bar{6}\ 2}$                \\
33:b-ac & p 21 a b   & p -2b 2a     & $\mathrm{p\ \bar{2}_{b}\ 2_{a}}$         & & 79      & p -6 2 m   & p -6 -2      & $\mathrm{p\ \bar{6}\ \bar{2}}$          \\
34      & p b 2 n    & p -2ab -2ab  & $\mathrm{p\ \bar{2}_{ab}\ \bar{2}_{ab}}$ & & 80      & p 6/m m m  & -p 6 2       & $\mathrm{\bar{p}\ 6\ 2}$                \\
\cline{1-4} \cline{6-9}
\end{tabular}
    }
    \caption{Table of the layer group numbers along with possible axis codes, Hermann–Mauguin symbols, Hall entries and Hall symbols. The table follows the format of table A1.4.2.7 in Ref.~\onlinecite{ITB}. The first row of each layer group number demonstrates the standard choice. All the choices are compatible with the symmetry diagrams in chapter 4 and table 1.2.6.1 of Ref.~\onlinecite{ITE}. The first letter 'p' and 'c' of the Hall symbols are in lowercase to represent layer groups.}
    \label{tab:hall-symbol-table}
\end{table*}

\providecommand{\noopsort}[1]{}\providecommand{\singleletter}[1]{#1}%


\begin{thebibliography}{52}%
\makeatletter
\providecommand \@ifxundefined [1]{%
 \@ifx{#1\undefined}
}%
\providecommand \@ifnum [1]{%
 \ifnum #1\expandafter \@firstoftwo
 \else \expandafter \@secondoftwo
 \fi
}%
\providecommand \@ifx [1]{%
 \ifx #1\expandafter \@firstoftwo
 \else \expandafter \@secondoftwo
 \fi
}%
\providecommand \natexlab [1]{#1}%
\providecommand \enquote  [1]{``#1''}%
\providecommand \bibnamefont  [1]{#1}%
\providecommand \bibfnamefont [1]{#1}%
\providecommand \citenamefont [1]{#1}%
\providecommand \href@noop [0]{\@secondoftwo}%
\providecommand \href [0]{\begingroup \@sanitize@url \@href}%
\providecommand \@href[1]{\@@startlink{#1}\@@href}%
\providecommand \@@href[1]{\endgroup#1\@@endlink}%
\providecommand \@sanitize@url [0]{\catcode `\\12\catcode `\$12\catcode
  `\&12\catcode `\#12\catcode `\^12\catcode `\_12\catcode `\%12\relax}%
\providecommand \@@startlink[1]{}%
\providecommand \@@endlink[0]{}%
\providecommand \url  [0]{\begingroup\@sanitize@url \@url }%
\providecommand \@url [1]{\endgroup\@href {#1}{\urlprefix }}%
\providecommand \urlprefix  [0]{URL }%
\providecommand \Eprint [0]{\href }%
\providecommand \doibase [0]{https://doi.org/}%
\providecommand \selectlanguage [0]{\@gobble}%
\providecommand \bibinfo  [0]{\@secondoftwo}%
\providecommand \bibfield  [0]{\@secondoftwo}%
\providecommand \translation [1]{[#1]}%
\providecommand \BibitemOpen [0]{}%
\providecommand \bibitemStop [0]{}%
\providecommand \bibitemNoStop [0]{.\EOS\space}%
\providecommand \EOS [0]{\spacefactor3000\relax}%
\providecommand \BibitemShut  [1]{\csname bibitem#1\endcsname}%
\let\auto@bib@innerbib\@empty
\bibitem [{\citenamefont {Hammermesh}\ and\ \citenamefont
  {Flammer}(1963)}]{hammermesh1963group}%
  \BibitemOpen
  \bibfield  {author} {\bibinfo {author} {\bibfnamefont {M.}~\bibnamefont
  {Hammermesh}}\ and\ \bibinfo {author} {\bibfnamefont {C.}~\bibnamefont
  {Flammer}},\ }\bibfield  {title} {\bibinfo {title} {Group theory and its
  application to physical problems},\ }\href@noop {} {\bibfield  {journal}
  {\bibinfo  {journal} {Physics Today}\ }\textbf {\bibinfo {volume} {16}},\
  \bibinfo {pages} {62} (\bibinfo {year} {1963})}\BibitemShut {NoStop}%
\bibitem [{\citenamefont {Wieder}\ \emph {et~al.}(2022)\citenamefont {Wieder},
  \citenamefont {Bradlyn}, \citenamefont {Cano}, \citenamefont {Wang},
  \citenamefont {Vergniory}, \citenamefont {Elcoro}, \citenamefont {Soluyanov},
  \citenamefont {Felser}, \citenamefont {Neupert}, \citenamefont {Regnault}
  \emph {et~al.}}]{wieder2022topological}%
  \BibitemOpen
  \bibfield  {author} {\bibinfo {author} {\bibfnamefont {B.~J.}\ \bibnamefont
  {Wieder}}, \bibinfo {author} {\bibfnamefont {B.}~\bibnamefont {Bradlyn}},
  \bibinfo {author} {\bibfnamefont {J.}~\bibnamefont {Cano}}, \bibinfo {author}
  {\bibfnamefont {Z.}~\bibnamefont {Wang}}, \bibinfo {author} {\bibfnamefont
  {M.~G.}\ \bibnamefont {Vergniory}}, \bibinfo {author} {\bibfnamefont
  {L.}~\bibnamefont {Elcoro}}, \bibinfo {author} {\bibfnamefont {A.~A.}\
  \bibnamefont {Soluyanov}}, \bibinfo {author} {\bibfnamefont {C.}~\bibnamefont
  {Felser}}, \bibinfo {author} {\bibfnamefont {T.}~\bibnamefont {Neupert}},
  \bibinfo {author} {\bibfnamefont {N.}~\bibnamefont {Regnault}}, \emph
  {et~al.},\ }\bibfield  {title} {\bibinfo {title} {Topological materials
  discovery from crystal symmetry},\ }\href@noop {} {\bibfield  {journal}
  {\bibinfo  {journal} {Nature Reviews Materials}\ }\textbf {\bibinfo {volume}
  {7}},\ \bibinfo {pages} {196} (\bibinfo {year} {2022})}\BibitemShut {NoStop}%
\bibitem [{\citenamefont {Boyd}(2020)}]{boyd2020nonlinear}%
  \BibitemOpen
  \bibfield  {author} {\bibinfo {author} {\bibfnamefont {R.~W.}\ \bibnamefont
  {Boyd}},\ }\href@noop {} {\emph {\bibinfo {title} {Nonlinear Optics}}}\
  (\bibinfo  {publisher} {Academic press},\ \bibinfo {year} {2020})\BibitemShut
  {NoStop}%
\bibitem [{\citenamefont {Briggs}\ \emph {et~al.}(2019)\citenamefont {Briggs},
  \citenamefont {Subramanian}, \citenamefont {Lin}, \citenamefont {Li},
  \citenamefont {Zhang}, \citenamefont {Zhang}, \citenamefont {Xiao},
  \citenamefont {Geohegan}, \citenamefont {Wallace}, \citenamefont {Chen} \emph
  {et~al.}}]{briggs2019roadmap}%
  \BibitemOpen
  \bibfield  {author} {\bibinfo {author} {\bibfnamefont {N.}~\bibnamefont
  {Briggs}}, \bibinfo {author} {\bibfnamefont {S.}~\bibnamefont {Subramanian}},
  \bibinfo {author} {\bibfnamefont {Z.}~\bibnamefont {Lin}}, \bibinfo {author}
  {\bibfnamefont {X.}~\bibnamefont {Li}}, \bibinfo {author} {\bibfnamefont
  {X.}~\bibnamefont {Zhang}}, \bibinfo {author} {\bibfnamefont
  {K.}~\bibnamefont {Zhang}}, \bibinfo {author} {\bibfnamefont
  {K.}~\bibnamefont {Xiao}}, \bibinfo {author} {\bibfnamefont {D.}~\bibnamefont
  {Geohegan}}, \bibinfo {author} {\bibfnamefont {R.}~\bibnamefont {Wallace}},
  \bibinfo {author} {\bibfnamefont {L.-Q.}\ \bibnamefont {Chen}}, \emph
  {et~al.},\ }\bibfield  {title} {\bibinfo {title} {A roadmap for electronic
  grade 2d materials},\ }\href@noop {} {\bibfield  {journal} {\bibinfo
  {journal} {2D Materials}\ }\textbf {\bibinfo {volume} {6}},\ \bibinfo {pages}
  {022001} (\bibinfo {year} {2019})}\BibitemShut {NoStop}%
\bibitem [{\citenamefont {Novoselov}\ \emph {et~al.}(2016)\citenamefont
  {Novoselov}, \citenamefont {Mishchenko}, \citenamefont {Carvalho},\ and\
  \citenamefont {Castro~Neto}}]{novoselov20162d}%
  \BibitemOpen
  \bibfield  {author} {\bibinfo {author} {\bibfnamefont {K.}~\bibnamefont
  {Novoselov}}, \bibinfo {author} {\bibfnamefont {o.~A.}\ \bibnamefont
  {Mishchenko}}, \bibinfo {author} {\bibfnamefont {o.~A.}\ \bibnamefont
  {Carvalho}},\ and\ \bibinfo {author} {\bibfnamefont {A.}~\bibnamefont
  {Castro~Neto}},\ }\bibfield  {title} {\bibinfo {title} {2d materials and van
  der waals heterostructures},\ }\href@noop {} {\bibfield  {journal} {\bibinfo
  {journal} {Science}\ }\textbf {\bibinfo {volume} {353}},\ \bibinfo {pages}
  {aac9439} (\bibinfo {year} {2016})}\BibitemShut {NoStop}%
\bibitem [{\citenamefont {Luo}\ \emph {et~al.}(2016)\citenamefont {Luo},
  \citenamefont {Liu},\ and\ \citenamefont {Wang}}]{luo2016recent}%
  \BibitemOpen
  \bibfield  {author} {\bibinfo {author} {\bibfnamefont {B.}~\bibnamefont
  {Luo}}, \bibinfo {author} {\bibfnamefont {G.}~\bibnamefont {Liu}},\ and\
  \bibinfo {author} {\bibfnamefont {L.}~\bibnamefont {Wang}},\ }\bibfield
  {title} {\bibinfo {title} {Recent advances in 2d materials for
  photocatalysis},\ }\href@noop {} {\bibfield  {journal} {\bibinfo  {journal}
  {Nanoscale}\ }\textbf {\bibinfo {volume} {8}},\ \bibinfo {pages} {6904}
  (\bibinfo {year} {2016})}\BibitemShut {NoStop}%
\bibitem [{\citenamefont {Peimyoo}\ \emph {et~al.}(2021)\citenamefont
  {Peimyoo}, \citenamefont {Deilmann}, \citenamefont {Withers}, \citenamefont
  {Escolar}, \citenamefont {Nutting}, \citenamefont {Taniguchi}, \citenamefont
  {Watanabe}, \citenamefont {Taghizadeh}, \citenamefont {Craciun},
  \citenamefont {Thygesen} \emph {et~al.}}]{peimyoo2021electrical}%
  \BibitemOpen
  \bibfield  {author} {\bibinfo {author} {\bibfnamefont {N.}~\bibnamefont
  {Peimyoo}}, \bibinfo {author} {\bibfnamefont {T.}~\bibnamefont {Deilmann}},
  \bibinfo {author} {\bibfnamefont {F.}~\bibnamefont {Withers}}, \bibinfo
  {author} {\bibfnamefont {J.}~\bibnamefont {Escolar}}, \bibinfo {author}
  {\bibfnamefont {D.}~\bibnamefont {Nutting}}, \bibinfo {author} {\bibfnamefont
  {T.}~\bibnamefont {Taniguchi}}, \bibinfo {author} {\bibfnamefont
  {K.}~\bibnamefont {Watanabe}}, \bibinfo {author} {\bibfnamefont
  {A.}~\bibnamefont {Taghizadeh}}, \bibinfo {author} {\bibfnamefont {M.~F.}\
  \bibnamefont {Craciun}}, \bibinfo {author} {\bibfnamefont {K.~S.}\
  \bibnamefont {Thygesen}}, \emph {et~al.},\ }\bibfield  {title} {\bibinfo
  {title} {Electrical tuning of optically active interlayer excitons in bilayer
  mos2},\ }\href@noop {} {\bibfield  {journal} {\bibinfo  {journal} {Nature
  Nanotechnology}\ }\textbf {\bibinfo {volume} {16}},\ \bibinfo {pages} {888}
  (\bibinfo {year} {2021})}\BibitemShut {NoStop}%
\bibitem [{\citenamefont {Zhao}\ \emph {et~al.}(2020)\citenamefont {Zhao},
  \citenamefont {Song}, \citenamefont {Wang}, \citenamefont {Riis-Jensen},
  \citenamefont {Fu}, \citenamefont {Deng}, \citenamefont {Wan}, \citenamefont
  {Kang}, \citenamefont {Ning}, \citenamefont {Dan} \emph
  {et~al.}}]{zhao2020engineering}%
  \BibitemOpen
  \bibfield  {author} {\bibinfo {author} {\bibfnamefont {X.}~\bibnamefont
  {Zhao}}, \bibinfo {author} {\bibfnamefont {P.}~\bibnamefont {Song}}, \bibinfo
  {author} {\bibfnamefont {C.}~\bibnamefont {Wang}}, \bibinfo {author}
  {\bibfnamefont {A.~C.}\ \bibnamefont {Riis-Jensen}}, \bibinfo {author}
  {\bibfnamefont {W.}~\bibnamefont {Fu}}, \bibinfo {author} {\bibfnamefont
  {Y.}~\bibnamefont {Deng}}, \bibinfo {author} {\bibfnamefont {D.}~\bibnamefont
  {Wan}}, \bibinfo {author} {\bibfnamefont {L.}~\bibnamefont {Kang}}, \bibinfo
  {author} {\bibfnamefont {S.}~\bibnamefont {Ning}}, \bibinfo {author}
  {\bibfnamefont {J.}~\bibnamefont {Dan}}, \emph {et~al.},\ }\bibfield  {title}
  {\bibinfo {title} {Engineering covalently bonded 2d layered materials by
  self-intercalation},\ }\href@noop {} {\bibfield  {journal} {\bibinfo
  {journal} {Nature}\ }\textbf {\bibinfo {volume} {581}},\ \bibinfo {pages}
  {171} (\bibinfo {year} {2020})}\BibitemShut {NoStop}%
\bibitem [{\citenamefont {Haastrup}\ \emph {et~al.}(2018)\citenamefont
  {Haastrup}, \citenamefont {Strange}, \citenamefont {Pandey}, \citenamefont
  {Deilmann}, \citenamefont {Schmidt}, \citenamefont {Hinsche}, \citenamefont
  {Gjerding}, \citenamefont {Torelli}, \citenamefont {Larsen}, \citenamefont
  {Riis-Jensen} \emph {et~al.}}]{haastrup2018computational}%
  \BibitemOpen
  \bibfield  {author} {\bibinfo {author} {\bibfnamefont {S.}~\bibnamefont
  {Haastrup}}, \bibinfo {author} {\bibfnamefont {M.}~\bibnamefont {Strange}},
  \bibinfo {author} {\bibfnamefont {M.}~\bibnamefont {Pandey}}, \bibinfo
  {author} {\bibfnamefont {T.}~\bibnamefont {Deilmann}}, \bibinfo {author}
  {\bibfnamefont {P.~S.}\ \bibnamefont {Schmidt}}, \bibinfo {author}
  {\bibfnamefont {N.~F.}\ \bibnamefont {Hinsche}}, \bibinfo {author}
  {\bibfnamefont {M.~N.}\ \bibnamefont {Gjerding}}, \bibinfo {author}
  {\bibfnamefont {D.}~\bibnamefont {Torelli}}, \bibinfo {author} {\bibfnamefont
  {P.~M.}\ \bibnamefont {Larsen}}, \bibinfo {author} {\bibfnamefont {A.~C.}\
  \bibnamefont {Riis-Jensen}}, \emph {et~al.},\ }\bibfield  {title} {\bibinfo
  {title} {The computational 2d materials database: high-throughput modeling
  and discovery of atomically thin crystals},\ }\href@noop {} {\bibfield
  {journal} {\bibinfo  {journal} {2D Materials}\ }\textbf {\bibinfo {volume}
  {5}},\ \bibinfo {pages} {042002} (\bibinfo {year} {2018})}\BibitemShut
  {NoStop}%
\bibitem [{\citenamefont {Gjerding}\ \emph {et~al.}(2021)\citenamefont
  {Gjerding}, \citenamefont {Taghizadeh}, \citenamefont {Rasmussen},
  \citenamefont {Ali}, \citenamefont {Bertoldo}, \citenamefont {Deilmann},
  \citenamefont {Kn{\o}sgaard}, \citenamefont {Kruse}, \citenamefont {Larsen},
  \citenamefont {Manti} \emph {et~al.}}]{gjerding2021recent}%
  \BibitemOpen
  \bibfield  {author} {\bibinfo {author} {\bibfnamefont {M.~N.}\ \bibnamefont
  {Gjerding}}, \bibinfo {author} {\bibfnamefont {A.}~\bibnamefont
  {Taghizadeh}}, \bibinfo {author} {\bibfnamefont {A.}~\bibnamefont
  {Rasmussen}}, \bibinfo {author} {\bibfnamefont {S.}~\bibnamefont {Ali}},
  \bibinfo {author} {\bibfnamefont {F.}~\bibnamefont {Bertoldo}}, \bibinfo
  {author} {\bibfnamefont {T.}~\bibnamefont {Deilmann}}, \bibinfo {author}
  {\bibfnamefont {N.~R.}\ \bibnamefont {Kn{\o}sgaard}}, \bibinfo {author}
  {\bibfnamefont {M.}~\bibnamefont {Kruse}}, \bibinfo {author} {\bibfnamefont
  {A.~H.}\ \bibnamefont {Larsen}}, \bibinfo {author} {\bibfnamefont
  {S.}~\bibnamefont {Manti}}, \emph {et~al.},\ }\bibfield  {title} {\bibinfo
  {title} {Recent progress of the computational 2d materials database (c2db)},\
  }\href@noop {} {\bibfield  {journal} {\bibinfo  {journal} {2D Materials}\
  }\textbf {\bibinfo {volume} {8}},\ \bibinfo {pages} {044002} (\bibinfo {year}
  {2021})}\BibitemShut {NoStop}%
\bibitem [{\citenamefont {Mounet}\ \emph {et~al.}(2018)\citenamefont {Mounet},
  \citenamefont {Gibertini}, \citenamefont {Schwaller}, \citenamefont {Campi},
  \citenamefont {Merkys}, \citenamefont {Marrazzo}, \citenamefont {Sohier},
  \citenamefont {Castelli}, \citenamefont {Cepellotti}, \citenamefont {Pizzi}
  \emph {et~al.}}]{mounet2018two}%
  \BibitemOpen
  \bibfield  {author} {\bibinfo {author} {\bibfnamefont {N.}~\bibnamefont
  {Mounet}}, \bibinfo {author} {\bibfnamefont {M.}~\bibnamefont {Gibertini}},
  \bibinfo {author} {\bibfnamefont {P.}~\bibnamefont {Schwaller}}, \bibinfo
  {author} {\bibfnamefont {D.}~\bibnamefont {Campi}}, \bibinfo {author}
  {\bibfnamefont {A.}~\bibnamefont {Merkys}}, \bibinfo {author} {\bibfnamefont
  {A.}~\bibnamefont {Marrazzo}}, \bibinfo {author} {\bibfnamefont
  {T.}~\bibnamefont {Sohier}}, \bibinfo {author} {\bibfnamefont {I.~E.}\
  \bibnamefont {Castelli}}, \bibinfo {author} {\bibfnamefont {A.}~\bibnamefont
  {Cepellotti}}, \bibinfo {author} {\bibfnamefont {G.}~\bibnamefont {Pizzi}},
  \emph {et~al.},\ }\bibfield  {title} {\bibinfo {title} {Two-dimensional
  materials from high-throughput computational exfoliation of experimentally
  known compounds},\ }\href@noop {} {\bibfield  {journal} {\bibinfo  {journal}
  {Nature nanotechnology}\ }\textbf {\bibinfo {volume} {13}},\ \bibinfo {pages}
  {246} (\bibinfo {year} {2018})}\BibitemShut {NoStop}%
\bibitem [{\citenamefont {Zhou}\ \emph {et~al.}(2019)\citenamefont {Zhou},
  \citenamefont {Shen}, \citenamefont {Costa}, \citenamefont {Persson},
  \citenamefont {Ong}, \citenamefont {Huck}, \citenamefont {Lu}, \citenamefont
  {Ma}, \citenamefont {Chen}, \citenamefont {Tang} \emph
  {et~al.}}]{zhou20192dmatpedia}%
  \BibitemOpen
  \bibfield  {author} {\bibinfo {author} {\bibfnamefont {J.}~\bibnamefont
  {Zhou}}, \bibinfo {author} {\bibfnamefont {L.}~\bibnamefont {Shen}}, \bibinfo
  {author} {\bibfnamefont {M.~D.}\ \bibnamefont {Costa}}, \bibinfo {author}
  {\bibfnamefont {K.~A.}\ \bibnamefont {Persson}}, \bibinfo {author}
  {\bibfnamefont {S.~P.}\ \bibnamefont {Ong}}, \bibinfo {author} {\bibfnamefont
  {P.}~\bibnamefont {Huck}}, \bibinfo {author} {\bibfnamefont {Y.}~\bibnamefont
  {Lu}}, \bibinfo {author} {\bibfnamefont {X.}~\bibnamefont {Ma}}, \bibinfo
  {author} {\bibfnamefont {Y.}~\bibnamefont {Chen}}, \bibinfo {author}
  {\bibfnamefont {H.}~\bibnamefont {Tang}}, \emph {et~al.},\ }\bibfield
  {title} {\bibinfo {title} {2dmatpedia, an open computational database of
  two-dimensional materials from top-down and bottom-up approaches},\
  }\href@noop {} {\bibfield  {journal} {\bibinfo  {journal} {Scientific data}\
  }\textbf {\bibinfo {volume} {6}},\ \bibinfo {pages} {86} (\bibinfo {year}
  {2019})}\BibitemShut {NoStop}%
\bibitem [{\citenamefont {Togo}\ and\ \citenamefont {Tanaka}(2018)}]{spglibv1}%
  \BibitemOpen
  \bibfield  {author} {\bibinfo {author} {\bibfnamefont {A.}~\bibnamefont
  {Togo}}\ and\ \bibinfo {author} {\bibfnamefont {I.}~\bibnamefont {Tanaka}},\
  }\href@noop {} {\bibinfo {title} {$\texttt{Spglib}$: a software library for
  crystal symmetry search}},\ \bibinfo {howpublished}
  {\url{https://github.com/spglib/spglib}} (\bibinfo {year} {2018})\BibitemShut
  {NoStop}%
\bibitem [{\citenamefont {Shinohara}\ \emph {et~al.}(2023)\citenamefont
  {Shinohara}, \citenamefont {Togo},\ and\ \citenamefont
  {Tanaka}}]{shinohara2023algorithms}%
  \BibitemOpen
  \bibfield  {author} {\bibinfo {author} {\bibfnamefont {K.}~\bibnamefont
  {Shinohara}}, \bibinfo {author} {\bibfnamefont {A.}~\bibnamefont {Togo}},\
  and\ \bibinfo {author} {\bibfnamefont {I.}~\bibnamefont {Tanaka}},\
  }\href@noop {} {\bibinfo {title} {Algorithms for magnetic symmetry operation
  search and identification of magnetic space group from magnetic crystal
  structure}} (\bibinfo {year} {2023})\BibitemShut {NoStop}%
\bibitem [{\citenamefont {Hanwell}\ \emph {et~al.}(2012)\citenamefont
  {Hanwell}, \citenamefont {Curtis}, \citenamefont {Lonie}, \citenamefont
  {Vandermeersch}, \citenamefont {Zurek},\ and\ \citenamefont
  {Hutchison}}]{Avogadro}%
  \BibitemOpen
  \bibfield  {author} {\bibinfo {author} {\bibfnamefont {M.~D.}\ \bibnamefont
  {Hanwell}}, \bibinfo {author} {\bibfnamefont {D.~E.}\ \bibnamefont {Curtis}},
  \bibinfo {author} {\bibfnamefont {D.~C.}\ \bibnamefont {Lonie}}, \bibinfo
  {author} {\bibfnamefont {T.}~\bibnamefont {Vandermeersch}}, \bibinfo {author}
  {\bibfnamefont {E.}~\bibnamefont {Zurek}},\ and\ \bibinfo {author}
  {\bibfnamefont {G.~R.}\ \bibnamefont {Hutchison}},\ }\bibfield  {title}
  {\bibinfo {title} {Avogadro: an advanced semantic chemical editor,
  visualization, and analysis platform},\ }\href
  {https://doi.org/10.1186/1758-2946-4-17} {\bibfield  {journal} {\bibinfo
  {journal} {Journal of Cheminformatics}\ }\textbf {\bibinfo {volume} {4}},\
  \bibinfo {pages} {17} (\bibinfo {year} {2012})}\BibitemShut {NoStop}%
\bibitem [{\citenamefont {Fredericks}\ \emph {et~al.}(2021)\citenamefont
  {Fredericks}, \citenamefont {Parrish}, \citenamefont {Sayre},\ and\
  \citenamefont {Zhu}}]{PyXtal}%
  \BibitemOpen
  \bibfield  {author} {\bibinfo {author} {\bibfnamefont {S.}~\bibnamefont
  {Fredericks}}, \bibinfo {author} {\bibfnamefont {K.}~\bibnamefont {Parrish}},
  \bibinfo {author} {\bibfnamefont {D.}~\bibnamefont {Sayre}},\ and\ \bibinfo
  {author} {\bibfnamefont {Q.}~\bibnamefont {Zhu}},\ }\bibfield  {title}
  {\bibinfo {title} {Pyxtal: A python library for crystal structure generation
  and symmetry analysis},\ }\href
  {https://doi.org/https://doi.org/10.1016/j.cpc.2020.107810} {\bibfield
  {journal} {\bibinfo  {journal} {Computer Physics Communications}\ }\textbf
  {\bibinfo {volume} {261}},\ \bibinfo {pages} {107810} (\bibinfo {year}
  {2021})}\BibitemShut {NoStop}%
\bibitem [{\citenamefont {Rahm}\ and\ \citenamefont
  {Erhart}(2020)}]{WulffPack}%
  \BibitemOpen
  \bibfield  {author} {\bibinfo {author} {\bibfnamefont {J.~M.}\ \bibnamefont
  {Rahm}}\ and\ \bibinfo {author} {\bibfnamefont {P.}~\bibnamefont {Erhart}},\
  }\bibfield  {title} {\bibinfo {title} {Wulffpack: A python package for wulff
  constructions},\ }\href {https://doi.org/10.21105/joss.01944} {\bibfield
  {journal} {\bibinfo  {journal} {Journal of Open Source Software}\ }\textbf
  {\bibinfo {volume} {5}},\ \bibinfo {pages} {1944} (\bibinfo {year}
  {2020})}\BibitemShut {NoStop}%
\bibitem [{\citenamefont {Kühne}\ \emph {et~al.}(2020)\citenamefont {Kühne},
  \citenamefont {Iannuzzi}, \citenamefont {Del~Ben}, \citenamefont {Rybkin},
  \citenamefont {Seewald}, \citenamefont {Stein}, \citenamefont {Laino},
  \citenamefont {Khaliullin}, \citenamefont {Schütt}, \citenamefont
  {Schiffmann}, \citenamefont {Golze}, \citenamefont {Wilhelm}, \citenamefont
  {Chulkov}, \citenamefont {Bani-Hashemian}, \citenamefont {Weber},
  \citenamefont {Borštnik}, \citenamefont {Taillefumier}, \citenamefont
  {Jakobovits}, \citenamefont {Lazzaro}, \citenamefont {Pabst}, \citenamefont
  {Müller}, \citenamefont {Schade}, \citenamefont {Guidon}, \citenamefont
  {Andermatt}, \citenamefont {Holmberg}, \citenamefont {Schenter},
  \citenamefont {Hehn}, \citenamefont {Bussy}, \citenamefont {Belleflamme},
  \citenamefont {Tabacchi}, \citenamefont {Glöß}, \citenamefont {Lass},
  \citenamefont {Bethune}, \citenamefont {Mundy}, \citenamefont {Plessl},
  \citenamefont {Watkins}, \citenamefont {VandeVondele}, \citenamefont
  {Krack},\ and\ \citenamefont {Hutter}}]{CP2K}%
  \BibitemOpen
  \bibfield  {author} {\bibinfo {author} {\bibfnamefont {T.~D.}\ \bibnamefont
  {Kühne}}, \bibinfo {author} {\bibfnamefont {M.}~\bibnamefont {Iannuzzi}},
  \bibinfo {author} {\bibfnamefont {M.}~\bibnamefont {Del~Ben}}, \bibinfo
  {author} {\bibfnamefont {V.~V.}\ \bibnamefont {Rybkin}}, \bibinfo {author}
  {\bibfnamefont {P.}~\bibnamefont {Seewald}}, \bibinfo {author} {\bibfnamefont
  {F.}~\bibnamefont {Stein}}, \bibinfo {author} {\bibfnamefont
  {T.}~\bibnamefont {Laino}}, \bibinfo {author} {\bibfnamefont {R.~Z.}\
  \bibnamefont {Khaliullin}}, \bibinfo {author} {\bibfnamefont
  {O.}~\bibnamefont {Schütt}}, \bibinfo {author} {\bibfnamefont
  {F.}~\bibnamefont {Schiffmann}}, \bibinfo {author} {\bibfnamefont
  {D.}~\bibnamefont {Golze}}, \bibinfo {author} {\bibfnamefont
  {J.}~\bibnamefont {Wilhelm}}, \bibinfo {author} {\bibfnamefont
  {S.}~\bibnamefont {Chulkov}}, \bibinfo {author} {\bibfnamefont {M.~H.}\
  \bibnamefont {Bani-Hashemian}}, \bibinfo {author} {\bibfnamefont
  {V.}~\bibnamefont {Weber}}, \bibinfo {author} {\bibfnamefont
  {U.}~\bibnamefont {Borštnik}}, \bibinfo {author} {\bibfnamefont
  {M.}~\bibnamefont {Taillefumier}}, \bibinfo {author} {\bibfnamefont {A.~S.}\
  \bibnamefont {Jakobovits}}, \bibinfo {author} {\bibfnamefont
  {A.}~\bibnamefont {Lazzaro}}, \bibinfo {author} {\bibfnamefont
  {H.}~\bibnamefont {Pabst}}, \bibinfo {author} {\bibfnamefont
  {T.}~\bibnamefont {Müller}}, \bibinfo {author} {\bibfnamefont
  {R.}~\bibnamefont {Schade}}, \bibinfo {author} {\bibfnamefont
  {M.}~\bibnamefont {Guidon}}, \bibinfo {author} {\bibfnamefont
  {S.}~\bibnamefont {Andermatt}}, \bibinfo {author} {\bibfnamefont
  {N.}~\bibnamefont {Holmberg}}, \bibinfo {author} {\bibfnamefont {G.~K.}\
  \bibnamefont {Schenter}}, \bibinfo {author} {\bibfnamefont {A.}~\bibnamefont
  {Hehn}}, \bibinfo {author} {\bibfnamefont {A.}~\bibnamefont {Bussy}},
  \bibinfo {author} {\bibfnamefont {F.}~\bibnamefont {Belleflamme}}, \bibinfo
  {author} {\bibfnamefont {G.}~\bibnamefont {Tabacchi}}, \bibinfo {author}
  {\bibfnamefont {A.}~\bibnamefont {Glöß}}, \bibinfo {author} {\bibfnamefont
  {M.}~\bibnamefont {Lass}}, \bibinfo {author} {\bibfnamefont {I.}~\bibnamefont
  {Bethune}}, \bibinfo {author} {\bibfnamefont {C.~J.}\ \bibnamefont {Mundy}},
  \bibinfo {author} {\bibfnamefont {C.}~\bibnamefont {Plessl}}, \bibinfo
  {author} {\bibfnamefont {M.}~\bibnamefont {Watkins}}, \bibinfo {author}
  {\bibfnamefont {J.}~\bibnamefont {VandeVondele}}, \bibinfo {author}
  {\bibfnamefont {M.}~\bibnamefont {Krack}},\ and\ \bibinfo {author}
  {\bibfnamefont {J.}~\bibnamefont {Hutter}},\ }\bibfield  {title} {\bibinfo
  {title} {{CP2K: An electronic structure and molecular dynamics software
  package - Quickstep: Efficient and accurate electronic structure
  calculations}},\ }\href {https://doi.org/10.1063/5.0007045} {\bibfield
  {journal} {\bibinfo  {journal} {The Journal of Chemical Physics}\ }\textbf
  {\bibinfo {volume} {152}},\ \bibinfo {pages} {194103} (\bibinfo {year}
  {2020})}\BibitemShut {NoStop}%
\bibitem [{\citenamefont {Herbst}\ \emph {et~al.}(2021)\citenamefont {Herbst},
  \citenamefont {Levitt},\ and\ \citenamefont {Cancès}}]{DFTKjcon}%
  \BibitemOpen
  \bibfield  {author} {\bibinfo {author} {\bibfnamefont {M.~F.}\ \bibnamefont
  {Herbst}}, \bibinfo {author} {\bibfnamefont {A.}~\bibnamefont {Levitt}},\
  and\ \bibinfo {author} {\bibfnamefont {E.}~\bibnamefont {Cancès}},\
  }\bibfield  {title} {\bibinfo {title} {Dftk: A julian approach for simulating
  electrons in solids},\ }\href {https://doi.org/10.21105/jcon.00069}
  {\bibfield  {journal} {\bibinfo  {journal} {Proc. JuliaCon Conf.}\ }\textbf
  {\bibinfo {volume} {3}},\ \bibinfo {pages} {69} (\bibinfo {year}
  {2021})}\BibitemShut {NoStop}%
\bibitem [{\citenamefont {Ångqvist}\ \emph {et~al.}(2019)\citenamefont
  {Ångqvist}, \citenamefont {Muñoz}, \citenamefont {Rahm}, \citenamefont
  {Fransson}, \citenamefont {Durniak}, \citenamefont {Rozyczko}, \citenamefont
  {Rod},\ and\ \citenamefont {Erhart}}]{ICET}%
  \BibitemOpen
  \bibfield  {author} {\bibinfo {author} {\bibfnamefont {M.}~\bibnamefont
  {Ångqvist}}, \bibinfo {author} {\bibfnamefont {W.~A.}\ \bibnamefont
  {Muñoz}}, \bibinfo {author} {\bibfnamefont {J.~M.}\ \bibnamefont {Rahm}},
  \bibinfo {author} {\bibfnamefont {E.}~\bibnamefont {Fransson}}, \bibinfo
  {author} {\bibfnamefont {C.}~\bibnamefont {Durniak}}, \bibinfo {author}
  {\bibfnamefont {P.}~\bibnamefont {Rozyczko}}, \bibinfo {author}
  {\bibfnamefont {T.~H.}\ \bibnamefont {Rod}},\ and\ \bibinfo {author}
  {\bibfnamefont {P.}~\bibnamefont {Erhart}},\ }\bibfield  {title} {\bibinfo
  {title} {Icet – a python library for constructing and sampling alloy
  cluster expansions},\ }\href
  {https://doi.org/https://doi.org/10.1002/adts.201900015} {\bibfield
  {journal} {\bibinfo  {journal} {Advanced Theory and Simulations}\ }\textbf
  {\bibinfo {volume} {2}},\ \bibinfo {pages} {1900015} (\bibinfo {year}
  {2019})}\BibitemShut {NoStop}%
\bibitem [{\citenamefont {Tancogne-Dejean}\ \emph {et~al.}(2020)\citenamefont
  {Tancogne-Dejean}, \citenamefont {Oliveira}, \citenamefont {Andrade},
  \citenamefont {Appel}, \citenamefont {Borca}, \citenamefont {Le~Breton},
  \citenamefont {Buchholz}, \citenamefont {Castro}, \citenamefont {Corni},
  \citenamefont {Correa}, \citenamefont {De~Giovannini}, \citenamefont
  {Delgado}, \citenamefont {Eich}, \citenamefont {Flick}, \citenamefont {Gil},
  \citenamefont {Gomez}, \citenamefont {Helbig}, \citenamefont {Hübener},
  \citenamefont {Jestädt}, \citenamefont {Jornet-Somoza}, \citenamefont
  {Larsen}, \citenamefont {Lebedeva}, \citenamefont {Lüders}, \citenamefont
  {Marques}, \citenamefont {Ohlmann}, \citenamefont {Pipolo}, \citenamefont
  {Rampp}, \citenamefont {Rozzi}, \citenamefont {Strubbe}, \citenamefont
  {Sato}, \citenamefont {Schäfer}, \citenamefont {Theophilou}, \citenamefont
  {Welden},\ and\ \citenamefont {Rubio}}]{Octopus}%
  \BibitemOpen
  \bibfield  {author} {\bibinfo {author} {\bibfnamefont {N.}~\bibnamefont
  {Tancogne-Dejean}}, \bibinfo {author} {\bibfnamefont {M.~J.~T.}\ \bibnamefont
  {Oliveira}}, \bibinfo {author} {\bibfnamefont {X.}~\bibnamefont {Andrade}},
  \bibinfo {author} {\bibfnamefont {H.}~\bibnamefont {Appel}}, \bibinfo
  {author} {\bibfnamefont {C.~H.}\ \bibnamefont {Borca}}, \bibinfo {author}
  {\bibfnamefont {G.}~\bibnamefont {Le~Breton}}, \bibinfo {author}
  {\bibfnamefont {F.}~\bibnamefont {Buchholz}}, \bibinfo {author}
  {\bibfnamefont {A.}~\bibnamefont {Castro}}, \bibinfo {author} {\bibfnamefont
  {S.}~\bibnamefont {Corni}}, \bibinfo {author} {\bibfnamefont {A.~A.}\
  \bibnamefont {Correa}}, \bibinfo {author} {\bibfnamefont {U.}~\bibnamefont
  {De~Giovannini}}, \bibinfo {author} {\bibfnamefont {A.}~\bibnamefont
  {Delgado}}, \bibinfo {author} {\bibfnamefont {F.~G.}\ \bibnamefont {Eich}},
  \bibinfo {author} {\bibfnamefont {J.}~\bibnamefont {Flick}}, \bibinfo
  {author} {\bibfnamefont {G.}~\bibnamefont {Gil}}, \bibinfo {author}
  {\bibfnamefont {A.}~\bibnamefont {Gomez}}, \bibinfo {author} {\bibfnamefont
  {N.}~\bibnamefont {Helbig}}, \bibinfo {author} {\bibfnamefont
  {H.}~\bibnamefont {Hübener}}, \bibinfo {author} {\bibfnamefont
  {R.}~\bibnamefont {Jestädt}}, \bibinfo {author} {\bibfnamefont
  {J.}~\bibnamefont {Jornet-Somoza}}, \bibinfo {author} {\bibfnamefont {A.~H.}\
  \bibnamefont {Larsen}}, \bibinfo {author} {\bibfnamefont {I.~V.}\
  \bibnamefont {Lebedeva}}, \bibinfo {author} {\bibfnamefont {M.}~\bibnamefont
  {Lüders}}, \bibinfo {author} {\bibfnamefont {M.~A.~L.}\ \bibnamefont
  {Marques}}, \bibinfo {author} {\bibfnamefont {S.~T.}\ \bibnamefont
  {Ohlmann}}, \bibinfo {author} {\bibfnamefont {S.}~\bibnamefont {Pipolo}},
  \bibinfo {author} {\bibfnamefont {M.}~\bibnamefont {Rampp}}, \bibinfo
  {author} {\bibfnamefont {C.~A.}\ \bibnamefont {Rozzi}}, \bibinfo {author}
  {\bibfnamefont {D.~A.}\ \bibnamefont {Strubbe}}, \bibinfo {author}
  {\bibfnamefont {S.~A.}\ \bibnamefont {Sato}}, \bibinfo {author}
  {\bibfnamefont {C.}~\bibnamefont {Schäfer}}, \bibinfo {author}
  {\bibfnamefont {I.}~\bibnamefont {Theophilou}}, \bibinfo {author}
  {\bibfnamefont {A.}~\bibnamefont {Welden}},\ and\ \bibinfo {author}
  {\bibfnamefont {A.}~\bibnamefont {Rubio}},\ }\bibfield  {title} {\bibinfo
  {title} {{Octopus, a computational framework for exploring light-driven
  phenomena and quantum dynamics in extended and finite systems}},\ }\href
  {https://doi.org/10.1063/1.5142502} {\bibfield  {journal} {\bibinfo
  {journal} {The Journal of Chemical Physics}\ }\textbf {\bibinfo {volume}
  {152}},\ \bibinfo {pages} {124119} (\bibinfo {year} {2020})}\BibitemShut
  {NoStop}%
\bibitem [{\citenamefont {Huber}\ \emph {et~al.}(2020)\citenamefont {Huber},
  \citenamefont {Zoupanos}, \citenamefont {Uhrin}, \citenamefont {Talirz},
  \citenamefont {Kahle}, \citenamefont {Häuselmann}, \citenamefont {Gresch},
  \citenamefont {Müller}, \citenamefont {Yakutovich}, \citenamefont
  {Andersen}, \citenamefont {Ramirez}, \citenamefont {Adorf}, \citenamefont
  {Gargiulo}, \citenamefont {Kumbhar}, \citenamefont {Passaro}, \citenamefont
  {Johnston}, \citenamefont {Merkys}, \citenamefont {Cepellotti}, \citenamefont
  {Mounet}, \citenamefont {Marzari}, \citenamefont {Kozinsky},\ and\
  \citenamefont {Pizzi}}]{AiiDA}%
  \BibitemOpen
  \bibfield  {author} {\bibinfo {author} {\bibfnamefont {S.~P.}\ \bibnamefont
  {Huber}}, \bibinfo {author} {\bibfnamefont {S.}~\bibnamefont {Zoupanos}},
  \bibinfo {author} {\bibfnamefont {M.}~\bibnamefont {Uhrin}}, \bibinfo
  {author} {\bibfnamefont {L.}~\bibnamefont {Talirz}}, \bibinfo {author}
  {\bibfnamefont {L.}~\bibnamefont {Kahle}}, \bibinfo {author} {\bibfnamefont
  {R.}~\bibnamefont {Häuselmann}}, \bibinfo {author} {\bibfnamefont
  {D.}~\bibnamefont {Gresch}}, \bibinfo {author} {\bibfnamefont
  {T.}~\bibnamefont {Müller}}, \bibinfo {author} {\bibfnamefont {A.~V.}\
  \bibnamefont {Yakutovich}}, \bibinfo {author} {\bibfnamefont {C.~W.}\
  \bibnamefont {Andersen}}, \bibinfo {author} {\bibfnamefont {F.~F.}\
  \bibnamefont {Ramirez}}, \bibinfo {author} {\bibfnamefont {C.~S.}\
  \bibnamefont {Adorf}}, \bibinfo {author} {\bibfnamefont {F.}~\bibnamefont
  {Gargiulo}}, \bibinfo {author} {\bibfnamefont {S.}~\bibnamefont {Kumbhar}},
  \bibinfo {author} {\bibfnamefont {E.}~\bibnamefont {Passaro}}, \bibinfo
  {author} {\bibfnamefont {C.}~\bibnamefont {Johnston}}, \bibinfo {author}
  {\bibfnamefont {A.}~\bibnamefont {Merkys}}, \bibinfo {author} {\bibfnamefont
  {A.}~\bibnamefont {Cepellotti}}, \bibinfo {author} {\bibfnamefont
  {N.}~\bibnamefont {Mounet}}, \bibinfo {author} {\bibfnamefont
  {N.}~\bibnamefont {Marzari}}, \bibinfo {author} {\bibfnamefont
  {B.}~\bibnamefont {Kozinsky}},\ and\ \bibinfo {author} {\bibfnamefont
  {G.}~\bibnamefont {Pizzi}},\ }\bibfield  {title} {\bibinfo {title} {{AiiDA
  1.0, a scalable computational infrastructure for automated reproducible
  workflows and data provenance}},\ }\href
  {https://doi.org/10.1038/s41597-020-00638-4} {\bibfield  {journal} {\bibinfo
  {journal} {Scientific Data}\ }\textbf {\bibinfo {volume} {7}},\ \bibinfo
  {pages} {300} (\bibinfo {year} {2020})}\BibitemShut {NoStop}%
\bibitem [{\citenamefont {Larsen}\ \emph {et~al.}(2017)\citenamefont {Larsen},
  \citenamefont {Mortensen}, \citenamefont {Blomqvist}, \citenamefont
  {Castelli}, \citenamefont {Christensen}, \citenamefont {Dułak},
  \citenamefont {Friis}, \citenamefont {Groves}, \citenamefont {Hammer},
  \citenamefont {Hargus}, \citenamefont {Hermes}, \citenamefont {Jennings},
  \citenamefont {Jensen}, \citenamefont {Kermode}, \citenamefont {Kitchin},
  \citenamefont {Kolsbjerg}, \citenamefont {Kubal}, \citenamefont {Kaasbjerg},
  \citenamefont {Lysgaard}, \citenamefont {Maronsson}, \citenamefont {Maxson},
  \citenamefont {Olsen}, \citenamefont {Pastewka}, \citenamefont {Peterson},
  \citenamefont {Rostgaard}, \citenamefont {Schiøtz}, \citenamefont {Schütt},
  \citenamefont {Strange}, \citenamefont {Thygesen}, \citenamefont {Vegge},
  \citenamefont {Vilhelmsen}, \citenamefont {Walter}, \citenamefont {Zeng},\
  and\ \citenamefont {Jacobsen}}]{ASE}%
  \BibitemOpen
  \bibfield  {author} {\bibinfo {author} {\bibfnamefont {A.~H.}\ \bibnamefont
  {Larsen}}, \bibinfo {author} {\bibfnamefont {J.~J.}\ \bibnamefont
  {Mortensen}}, \bibinfo {author} {\bibfnamefont {J.}~\bibnamefont
  {Blomqvist}}, \bibinfo {author} {\bibfnamefont {I.~E.}\ \bibnamefont
  {Castelli}}, \bibinfo {author} {\bibfnamefont {R.}~\bibnamefont
  {Christensen}}, \bibinfo {author} {\bibfnamefont {M.}~\bibnamefont {Dułak}},
  \bibinfo {author} {\bibfnamefont {J.}~\bibnamefont {Friis}}, \bibinfo
  {author} {\bibfnamefont {M.~N.}\ \bibnamefont {Groves}}, \bibinfo {author}
  {\bibfnamefont {B.}~\bibnamefont {Hammer}}, \bibinfo {author} {\bibfnamefont
  {C.}~\bibnamefont {Hargus}}, \bibinfo {author} {\bibfnamefont {E.~D.}\
  \bibnamefont {Hermes}}, \bibinfo {author} {\bibfnamefont {P.~C.}\
  \bibnamefont {Jennings}}, \bibinfo {author} {\bibfnamefont {P.~B.}\
  \bibnamefont {Jensen}}, \bibinfo {author} {\bibfnamefont {J.}~\bibnamefont
  {Kermode}}, \bibinfo {author} {\bibfnamefont {J.~R.}\ \bibnamefont
  {Kitchin}}, \bibinfo {author} {\bibfnamefont {E.~L.}\ \bibnamefont
  {Kolsbjerg}}, \bibinfo {author} {\bibfnamefont {J.}~\bibnamefont {Kubal}},
  \bibinfo {author} {\bibfnamefont {K.}~\bibnamefont {Kaasbjerg}}, \bibinfo
  {author} {\bibfnamefont {S.}~\bibnamefont {Lysgaard}}, \bibinfo {author}
  {\bibfnamefont {J.~B.}\ \bibnamefont {Maronsson}}, \bibinfo {author}
  {\bibfnamefont {T.}~\bibnamefont {Maxson}}, \bibinfo {author} {\bibfnamefont
  {T.}~\bibnamefont {Olsen}}, \bibinfo {author} {\bibfnamefont
  {L.}~\bibnamefont {Pastewka}}, \bibinfo {author} {\bibfnamefont
  {A.}~\bibnamefont {Peterson}}, \bibinfo {author} {\bibfnamefont
  {C.}~\bibnamefont {Rostgaard}}, \bibinfo {author} {\bibfnamefont
  {J.}~\bibnamefont {Schiøtz}}, \bibinfo {author} {\bibfnamefont
  {O.}~\bibnamefont {Schütt}}, \bibinfo {author} {\bibfnamefont
  {M.}~\bibnamefont {Strange}}, \bibinfo {author} {\bibfnamefont {K.~S.}\
  \bibnamefont {Thygesen}}, \bibinfo {author} {\bibfnamefont {T.}~\bibnamefont
  {Vegge}}, \bibinfo {author} {\bibfnamefont {L.}~\bibnamefont {Vilhelmsen}},
  \bibinfo {author} {\bibfnamefont {M.}~\bibnamefont {Walter}}, \bibinfo
  {author} {\bibfnamefont {Z.}~\bibnamefont {Zeng}},\ and\ \bibinfo {author}
  {\bibfnamefont {K.~W.}\ \bibnamefont {Jacobsen}},\ }\bibfield  {title}
  {\bibinfo {title} {The atomic simulation environment—a python library for
  working with atoms},\ }\href {https://doi.org/10.1088/1361-648X/aa680e}
  {\bibfield  {journal} {\bibinfo  {journal} {Journal of Physics: Condensed
  Matter}\ }\textbf {\bibinfo {volume} {29}},\ \bibinfo {pages} {273002}
  (\bibinfo {year} {2017})}\BibitemShut {NoStop}%
\bibitem [{\citenamefont {Ward}\ \emph {et~al.}(2018)\citenamefont {Ward},
  \citenamefont {Dunn}, \citenamefont {Faghaninia}, \citenamefont {Zimmermann},
  \citenamefont {Bajaj}, \citenamefont {Wang}, \citenamefont {Montoya},
  \citenamefont {Chen}, \citenamefont {Bystrom}, \citenamefont {Dylla},
  \citenamefont {Chard}, \citenamefont {Asta}, \citenamefont {Persson},
  \citenamefont {Snyder}, \citenamefont {Foster},\ and\ \citenamefont
  {Jain}}]{Matminer}%
  \BibitemOpen
  \bibfield  {author} {\bibinfo {author} {\bibfnamefont {L.}~\bibnamefont
  {Ward}}, \bibinfo {author} {\bibfnamefont {A.}~\bibnamefont {Dunn}}, \bibinfo
  {author} {\bibfnamefont {A.}~\bibnamefont {Faghaninia}}, \bibinfo {author}
  {\bibfnamefont {N.~E.}\ \bibnamefont {Zimmermann}}, \bibinfo {author}
  {\bibfnamefont {S.}~\bibnamefont {Bajaj}}, \bibinfo {author} {\bibfnamefont
  {Q.}~\bibnamefont {Wang}}, \bibinfo {author} {\bibfnamefont {J.}~\bibnamefont
  {Montoya}}, \bibinfo {author} {\bibfnamefont {J.}~\bibnamefont {Chen}},
  \bibinfo {author} {\bibfnamefont {K.}~\bibnamefont {Bystrom}}, \bibinfo
  {author} {\bibfnamefont {M.}~\bibnamefont {Dylla}}, \bibinfo {author}
  {\bibfnamefont {K.}~\bibnamefont {Chard}}, \bibinfo {author} {\bibfnamefont
  {M.}~\bibnamefont {Asta}}, \bibinfo {author} {\bibfnamefont {K.~A.}\
  \bibnamefont {Persson}}, \bibinfo {author} {\bibfnamefont {G.~J.}\
  \bibnamefont {Snyder}}, \bibinfo {author} {\bibfnamefont {I.}~\bibnamefont
  {Foster}},\ and\ \bibinfo {author} {\bibfnamefont {A.}~\bibnamefont {Jain}},\
  }\bibfield  {title} {\bibinfo {title} {Matminer: An open source toolkit for
  materials data mining},\ }\href
  {https://doi.org/https://doi.org/10.1016/j.commatsci.2018.05.018} {\bibfield
  {journal} {\bibinfo  {journal} {Computational Materials Science}\ }\textbf
  {\bibinfo {volume} {152}},\ \bibinfo {pages} {60} (\bibinfo {year}
  {2018})}\BibitemShut {NoStop}%
\bibitem [{\citenamefont {Krogel}(2016)}]{Nexus}%
  \BibitemOpen
  \bibfield  {author} {\bibinfo {author} {\bibfnamefont {J.~T.}\ \bibnamefont
  {Krogel}},\ }\bibfield  {title} {\bibinfo {title} {Nexus: A modular workflow
  management system for quantum simulation codes},\ }\href
  {https://doi.org/https://doi.org/10.1016/j.cpc.2015.08.012} {\bibfield
  {journal} {\bibinfo  {journal} {Computer Physics Communications}\ }\textbf
  {\bibinfo {volume} {198}},\ \bibinfo {pages} {154} (\bibinfo {year}
  {2016})}\BibitemShut {NoStop}%
\bibitem [{\citenamefont {Ong}\ \emph {et~al.}(2013)\citenamefont {Ong},
  \citenamefont {Richards}, \citenamefont {Jain}, \citenamefont {Hautier},
  \citenamefont {Kocher}, \citenamefont {Cholia}, \citenamefont {Gunter},
  \citenamefont {Chevrier}, \citenamefont {Persson},\ and\ \citenamefont
  {Ceder}}]{pymatgen}%
  \BibitemOpen
  \bibfield  {author} {\bibinfo {author} {\bibfnamefont {S.~P.}\ \bibnamefont
  {Ong}}, \bibinfo {author} {\bibfnamefont {W.~D.}\ \bibnamefont {Richards}},
  \bibinfo {author} {\bibfnamefont {A.}~\bibnamefont {Jain}}, \bibinfo {author}
  {\bibfnamefont {G.}~\bibnamefont {Hautier}}, \bibinfo {author} {\bibfnamefont
  {M.}~\bibnamefont {Kocher}}, \bibinfo {author} {\bibfnamefont
  {S.}~\bibnamefont {Cholia}}, \bibinfo {author} {\bibfnamefont
  {D.}~\bibnamefont {Gunter}}, \bibinfo {author} {\bibfnamefont {V.~L.}\
  \bibnamefont {Chevrier}}, \bibinfo {author} {\bibfnamefont {K.~A.}\
  \bibnamefont {Persson}},\ and\ \bibinfo {author} {\bibfnamefont
  {G.}~\bibnamefont {Ceder}},\ }\bibfield  {title} {\bibinfo {title} {Python
  materials genomics (pymatgen): A robust, open-source python library for
  materials analysis},\ }\href
  {https://doi.org/https://doi.org/10.1016/j.commatsci.2012.10.028} {\bibfield
  {journal} {\bibinfo  {journal} {Computational Materials Science}\ }\textbf
  {\bibinfo {volume} {68}},\ \bibinfo {pages} {314} (\bibinfo {year}
  {2013})}\BibitemShut {NoStop}%
\bibitem [{\citenamefont {Eriksson}\ \emph {et~al.}(2019)\citenamefont
  {Eriksson}, \citenamefont {Fransson},\ and\ \citenamefont
  {Erhart}}]{hiphive}%
  \BibitemOpen
  \bibfield  {author} {\bibinfo {author} {\bibfnamefont {F.}~\bibnamefont
  {Eriksson}}, \bibinfo {author} {\bibfnamefont {E.}~\bibnamefont {Fransson}},\
  and\ \bibinfo {author} {\bibfnamefont {P.}~\bibnamefont {Erhart}},\
  }\bibfield  {title} {\bibinfo {title} {The hiphive package for the extraction
  of high-order force constants by machine learning},\ }\href
  {https://doi.org/https://doi.org/10.1002/adts.201800184} {\bibfield
  {journal} {\bibinfo  {journal} {Advanced Theory and Simulations}\ }\textbf
  {\bibinfo {volume} {2}},\ \bibinfo {pages} {1800184} (\bibinfo {year}
  {2019})}\BibitemShut {NoStop}%
\bibitem [{\citenamefont {Togo}\ \emph {et~al.}(2023)\citenamefont {Togo},
  \citenamefont {Chaput}, \citenamefont {Tadano},\ and\ \citenamefont
  {Tanaka}}]{Phonopy}%
  \BibitemOpen
  \bibfield  {author} {\bibinfo {author} {\bibfnamefont {A.}~\bibnamefont
  {Togo}}, \bibinfo {author} {\bibfnamefont {L.}~\bibnamefont {Chaput}},
  \bibinfo {author} {\bibfnamefont {T.}~\bibnamefont {Tadano}},\ and\ \bibinfo
  {author} {\bibfnamefont {I.}~\bibnamefont {Tanaka}},\ }\bibfield  {title}
  {\bibinfo {title} {Implementation strategies in phonopy and phono3py},\
  }\href {https://doi.org/10.1088/1361-648X/acd831} {\bibfield  {journal}
  {\bibinfo  {journal} {Journal of Physics: Condensed Matter}\ }\textbf
  {\bibinfo {volume} {35}},\ \bibinfo {pages} {353001} (\bibinfo {year}
  {2023})}\BibitemShut {NoStop}%
\bibitem [{\citenamefont {Hinuma}\ \emph {et~al.}(2017)\citenamefont {Hinuma},
  \citenamefont {Pizzi}, \citenamefont {Kumagai}, \citenamefont {Oba},\ and\
  \citenamefont {Tanaka}}]{SeeKpath}%
  \BibitemOpen
  \bibfield  {author} {\bibinfo {author} {\bibfnamefont {Y.}~\bibnamefont
  {Hinuma}}, \bibinfo {author} {\bibfnamefont {G.}~\bibnamefont {Pizzi}},
  \bibinfo {author} {\bibfnamefont {Y.}~\bibnamefont {Kumagai}}, \bibinfo
  {author} {\bibfnamefont {F.}~\bibnamefont {Oba}},\ and\ \bibinfo {author}
  {\bibfnamefont {I.}~\bibnamefont {Tanaka}},\ }\bibfield  {title} {\bibinfo
  {title} {Band structure diagram paths based on crystallography},\ }\href
  {https://doi.org/https://doi.org/10.1016/j.commatsci.2016.10.015} {\bibfield
  {journal} {\bibinfo  {journal} {Computational Materials Science}\ }\textbf
  {\bibinfo {volume} {128}},\ \bibinfo {pages} {140} (\bibinfo {year}
  {2017})}\BibitemShut {NoStop}%
\bibitem [{\citenamefont {Choudhary}\ \emph {et~al.}(2020)\citenamefont
  {Choudhary}, \citenamefont {Garrity}, \citenamefont {Reid}, \citenamefont
  {DeCost}, \citenamefont {Biacchi}, \citenamefont {Hight~Walker},
  \citenamefont {Trautt}, \citenamefont {Hattrick-Simpers}, \citenamefont
  {Kusne}, \citenamefont {Centrone}, \citenamefont {Davydov}, \citenamefont
  {Jiang}, \citenamefont {Pachter}, \citenamefont {Cheon}, \citenamefont
  {Reed}, \citenamefont {Agrawal}, \citenamefont {Qian}, \citenamefont
  {Sharma}, \citenamefont {Zhuang}, \citenamefont {Kalinin}, \citenamefont
  {Sumpter}, \citenamefont {Pilania}, \citenamefont {Acar}, \citenamefont
  {Mandal}, \citenamefont {Haule}, \citenamefont {Vanderbilt}, \citenamefont
  {Rabe},\ and\ \citenamefont {Tavazza}}]{JARVIS}%
  \BibitemOpen
  \bibfield  {author} {\bibinfo {author} {\bibfnamefont {K.}~\bibnamefont
  {Choudhary}}, \bibinfo {author} {\bibfnamefont {K.~F.}\ \bibnamefont
  {Garrity}}, \bibinfo {author} {\bibfnamefont {A.~C.~E.}\ \bibnamefont
  {Reid}}, \bibinfo {author} {\bibfnamefont {B.}~\bibnamefont {DeCost}},
  \bibinfo {author} {\bibfnamefont {A.~J.}\ \bibnamefont {Biacchi}}, \bibinfo
  {author} {\bibfnamefont {A.~R.}\ \bibnamefont {Hight~Walker}}, \bibinfo
  {author} {\bibfnamefont {Z.}~\bibnamefont {Trautt}}, \bibinfo {author}
  {\bibfnamefont {J.}~\bibnamefont {Hattrick-Simpers}}, \bibinfo {author}
  {\bibfnamefont {A.~G.}\ \bibnamefont {Kusne}}, \bibinfo {author}
  {\bibfnamefont {A.}~\bibnamefont {Centrone}}, \bibinfo {author}
  {\bibfnamefont {A.}~\bibnamefont {Davydov}}, \bibinfo {author} {\bibfnamefont
  {J.}~\bibnamefont {Jiang}}, \bibinfo {author} {\bibfnamefont
  {R.}~\bibnamefont {Pachter}}, \bibinfo {author} {\bibfnamefont
  {G.}~\bibnamefont {Cheon}}, \bibinfo {author} {\bibfnamefont
  {E.}~\bibnamefont {Reed}}, \bibinfo {author} {\bibfnamefont {A.}~\bibnamefont
  {Agrawal}}, \bibinfo {author} {\bibfnamefont {X.}~\bibnamefont {Qian}},
  \bibinfo {author} {\bibfnamefont {V.}~\bibnamefont {Sharma}}, \bibinfo
  {author} {\bibfnamefont {H.}~\bibnamefont {Zhuang}}, \bibinfo {author}
  {\bibfnamefont {S.~V.}\ \bibnamefont {Kalinin}}, \bibinfo {author}
  {\bibfnamefont {B.~G.}\ \bibnamefont {Sumpter}}, \bibinfo {author}
  {\bibfnamefont {G.}~\bibnamefont {Pilania}}, \bibinfo {author} {\bibfnamefont
  {P.}~\bibnamefont {Acar}}, \bibinfo {author} {\bibfnamefont {S.}~\bibnamefont
  {Mandal}}, \bibinfo {author} {\bibfnamefont {K.}~\bibnamefont {Haule}},
  \bibinfo {author} {\bibfnamefont {D.}~\bibnamefont {Vanderbilt}}, \bibinfo
  {author} {\bibfnamefont {K.}~\bibnamefont {Rabe}},\ and\ \bibinfo {author}
  {\bibfnamefont {F.}~\bibnamefont {Tavazza}},\ }\bibfield  {title} {\bibinfo
  {title} {The joint automated repository for various integrated simulations
  (jarvis) for data-driven materials design},\ }\href
  {https://doi.org/10.1038/s41524-020-00440-1} {\bibfield  {journal} {\bibinfo
  {journal} {npj Computational Materials}\ }\textbf {\bibinfo {volume} {6}},\
  \bibinfo {pages} {173} (\bibinfo {year} {2020})}\BibitemShut {NoStop}%
\bibitem [{\citenamefont {Jain}\ \emph {et~al.}(2013)\citenamefont {Jain},
  \citenamefont {Ong}, \citenamefont {Hautier}, \citenamefont {Chen},
  \citenamefont {Richards}, \citenamefont {Dacek}, \citenamefont {Cholia},
  \citenamefont {Gunter}, \citenamefont {Skinner}, \citenamefont {Ceder},\ and\
  \citenamefont {Persson}}]{MaterialsProject}%
  \BibitemOpen
  \bibfield  {author} {\bibinfo {author} {\bibfnamefont {A.}~\bibnamefont
  {Jain}}, \bibinfo {author} {\bibfnamefont {S.~P.}\ \bibnamefont {Ong}},
  \bibinfo {author} {\bibfnamefont {G.}~\bibnamefont {Hautier}}, \bibinfo
  {author} {\bibfnamefont {W.}~\bibnamefont {Chen}}, \bibinfo {author}
  {\bibfnamefont {W.~D.}\ \bibnamefont {Richards}}, \bibinfo {author}
  {\bibfnamefont {S.}~\bibnamefont {Dacek}}, \bibinfo {author} {\bibfnamefont
  {S.}~\bibnamefont {Cholia}}, \bibinfo {author} {\bibfnamefont
  {D.}~\bibnamefont {Gunter}}, \bibinfo {author} {\bibfnamefont
  {D.}~\bibnamefont {Skinner}}, \bibinfo {author} {\bibfnamefont
  {G.}~\bibnamefont {Ceder}},\ and\ \bibinfo {author} {\bibfnamefont {K.~A.}\
  \bibnamefont {Persson}},\ }\bibfield  {title} {\bibinfo {title} {{Commentary:
  The Materials Project: A materials genome approach to accelerating materials
  innovation}},\ }\href {https://doi.org/10.1063/1.4812323} {\bibfield
  {journal} {\bibinfo  {journal} {APL Materials}\ }\textbf {\bibinfo {volume}
  {1}},\ \bibinfo {pages} {011002} (\bibinfo {year} {2013})}\BibitemShut
  {NoStop}%
\bibitem [{\citenamefont {Kirklin}\ \emph {et~al.}(2015)\citenamefont
  {Kirklin}, \citenamefont {Saal}, \citenamefont {Meredig}, \citenamefont
  {Thompson}, \citenamefont {Doak}, \citenamefont {Aykol}, \citenamefont
  {R{\"u}hl},\ and\ \citenamefont {Wolverton}}]{OQMD}%
  \BibitemOpen
  \bibfield  {author} {\bibinfo {author} {\bibfnamefont {S.}~\bibnamefont
  {Kirklin}}, \bibinfo {author} {\bibfnamefont {J.~E.}\ \bibnamefont {Saal}},
  \bibinfo {author} {\bibfnamefont {B.}~\bibnamefont {Meredig}}, \bibinfo
  {author} {\bibfnamefont {A.}~\bibnamefont {Thompson}}, \bibinfo {author}
  {\bibfnamefont {J.~W.}\ \bibnamefont {Doak}}, \bibinfo {author}
  {\bibfnamefont {M.}~\bibnamefont {Aykol}}, \bibinfo {author} {\bibfnamefont
  {S.}~\bibnamefont {R{\"u}hl}},\ and\ \bibinfo {author} {\bibfnamefont
  {C.}~\bibnamefont {Wolverton}},\ }\bibfield  {title} {\bibinfo {title} {The
  open quantum materials database (oqmd): assessing the accuracy of dft
  formation energies},\ }\href {https://doi.org/10.1038/npjcompumats.2015.10}
  {\bibfield  {journal} {\bibinfo  {journal} {npj Computational Materials}\
  }\textbf {\bibinfo {volume} {1}},\ \bibinfo {pages} {15010} (\bibinfo {year}
  {2015})}\BibitemShut {NoStop}%
\bibitem [{\citenamefont {Kopsky}\ and\ \citenamefont {Litvin}(2010)}]{ITE}%
  \BibitemOpen
  \bibfield  {author} {\bibinfo {author} {\bibfnamefont {V.}~\bibnamefont
  {Kopsky}}\ and\ \bibinfo {author} {\bibfnamefont {D.~B.}\ \bibnamefont
  {Litvin}},\ }\href@noop {} {\emph {\bibinfo {title} {International tables for
  crystallography, Vol. E, Subperiodic groups}}}\ (\bibinfo  {publisher} {John
  Wiley \& Sons, Ltd},\ \bibinfo {year} {2010})\BibitemShut {NoStop}%
\bibitem [{\citenamefont {Aroyo}(2016)}]{ITA}%
  \BibitemOpen
  \bibfield  {author} {\bibinfo {author} {\bibfnamefont {M.~I.}\ \bibnamefont
  {Aroyo}},\ }\href@noop {} {\emph {\bibinfo {title} {International tables for
  crystallography, Vol. A, Space-group symmetry}}}\ (\bibinfo  {publisher}
  {John Wiley \& Sons, Ltd},\ \bibinfo {year} {2016})\BibitemShut {NoStop}%
\bibitem [{\citenamefont {Delaunay}(1933)}]{Delaunay1933}%
  \BibitemOpen
  \bibfield  {author} {\bibinfo {author} {\bibfnamefont {B.}~\bibnamefont
  {Delaunay}},\ }\bibfield  {title} {\bibinfo {title} {Neue darstellung der
  geometrischen kristallographie},\ }\href
  {https://doi.org/doi:10.1524/zkri.1933.84.1.109} {\bibfield  {journal}
  {\bibinfo  {journal} {Zeitschrift für Kristallographie - Crystalline
  Materials}\ }\textbf {\bibinfo {volume} {84}},\ \bibinfo {pages} {109}
  (\bibinfo {year} {1933})}\BibitemShut {NoStop}%
\bibitem [{\citenamefont {Burzlaff}\ and\ \citenamefont
  {Zimmermann}(1985)}]{BurzlaffZimmermann1985}%
  \BibitemOpen
  \bibfield  {author} {\bibinfo {author} {\bibfnamefont {H.}~\bibnamefont
  {Burzlaff}}\ and\ \bibinfo {author} {\bibfnamefont {H.}~\bibnamefont
  {Zimmermann}},\ }\bibfield  {title} {\bibinfo {title} {On the metrical
  properties of lattices},\ }\href
  {https://doi.org/doi:10.1524/zkri.1985.170.14.247} {\bibfield  {journal}
  {\bibinfo  {journal} {Zeitschrift für Kristallographie - Crystalline
  Materials}\ }\textbf {\bibinfo {volume} {170}},\ \bibinfo {pages} {247}
  (\bibinfo {year} {1985})}\BibitemShut {NoStop}%
\bibitem [{\citenamefont {Hall}(1981)}]{Hall:a19707}%
  \BibitemOpen
  \bibfield  {author} {\bibinfo {author} {\bibfnamefont {S.~R.}\ \bibnamefont
  {Hall}},\ }\bibfield  {title} {\bibinfo {title} {{Space-group notation with
  an explicit origin}},\ }\href {https://doi.org/10.1107/S0567739481001228}
  {\bibfield  {journal} {\bibinfo  {journal} {Acta Crystallographica Section
  A}\ }\textbf {\bibinfo {volume} {37}},\ \bibinfo {pages} {517} (\bibinfo
  {year} {1981})}\BibitemShut {NoStop}%
\bibitem [{\citenamefont {Shmueli}\ \emph {et~al.}(2010)\citenamefont
  {Shmueli}, \citenamefont {Hall},\ and\ \citenamefont
  {Grosse-Kunstleve}}]{ITB}%
  \BibitemOpen
  \bibfield  {author} {\bibinfo {author} {\bibfnamefont {U.}~\bibnamefont
  {Shmueli}}, \bibinfo {author} {\bibfnamefont {S.~R.}\ \bibnamefont {Hall}},\
  and\ \bibinfo {author} {\bibfnamefont {R.~W.}\ \bibnamefont
  {Grosse-Kunstleve}},\ }\bibinfo {title} {Symmetry in reciprocal space},\ in\
  \href {https://doi.org/https://doi.org/10.1107/97809553602060000761} {\emph
  {\bibinfo {booktitle} {International Tables for Crystallography, Vol. B,
  Reciprocal space}}}\ (\bibinfo  {publisher} {John Wiley \& Sons, Ltd},\
  \bibinfo {year} {2010})\ Chap.\ \bibinfo {chapter} {1.4}, pp.\ \bibinfo
  {pages} {114--174}\BibitemShut {NoStop}%
\bibitem [{\citenamefont {Grosse-Kunstleve}(1999)}]{Grosse-Kunstleve:au0146}%
  \BibitemOpen
  \bibfield  {author} {\bibinfo {author} {\bibfnamefont {R.~W.}\ \bibnamefont
  {Grosse-Kunstleve}},\ }\bibfield  {title} {\bibinfo {title} {{Algorithms for
  deriving crystallographic space-group information}},\ }\href
  {https://doi.org/10.1107/S0108767398010186} {\bibfield  {journal} {\bibinfo
  {journal} {Acta Crystallographica Section A}\ }\textbf {\bibinfo {volume}
  {55}},\ \bibinfo {pages} {383} (\bibinfo {year} {1999})}\BibitemShut
  {NoStop}%
\bibitem [{\citenamefont {Lyngby}\ and\ \citenamefont
  {Thygesen}(2022)}]{Lyngby2022}%
  \BibitemOpen
  \bibfield  {author} {\bibinfo {author} {\bibfnamefont {P.}~\bibnamefont
  {Lyngby}}\ and\ \bibinfo {author} {\bibfnamefont {K.~S.}\ \bibnamefont
  {Thygesen}},\ }\bibfield  {title} {\bibinfo {title} {Data-driven discovery of
  2d materials by deep generative models},\ }\href
  {https://doi.org/10.1038/s41524-022-00923-3} {\bibfield  {journal} {\bibinfo
  {journal} {npj Computational Materials}\ }\textbf {\bibinfo {volume} {8}},\
  \bibinfo {pages} {232} (\bibinfo {year} {2022})}\BibitemShut {NoStop}%
\bibitem [{\citenamefont {Kopsk{\'{y}}}(1989)}]{Kopsk&yacute;:bx0374}%
  \BibitemOpen
  \bibfield  {author} {\bibinfo {author} {\bibfnamefont {V.}~\bibnamefont
  {Kopsk{\'{y}}}},\ }\bibfield  {title} {\bibinfo {title} {{Subperiodic classes
  of reducible space groups}},\ }\href
  {https://doi.org/10.1107/S0108767389008172} {\bibfield  {journal} {\bibinfo
  {journal} {Acta Crystallographica Section A}\ }\textbf {\bibinfo {volume}
  {45}},\ \bibinfo {pages} {815} (\bibinfo {year} {1989})}\BibitemShut
  {NoStop}%
\bibitem [{\citenamefont {Kopsk{\'{y}}}(1993)}]{Kopsk&yacute;:bx0469}%
  \BibitemOpen
  \bibfield  {author} {\bibinfo {author} {\bibfnamefont {V.}~\bibnamefont
  {Kopsk{\'{y}}}},\ }\bibfield  {title} {\bibinfo {title} {{Layer and rod
  classes of reducible space groups. I. {\it Z}-decomposable cases}},\ }\href
  {https://doi.org/10.1107/S0108767392006585} {\bibfield  {journal} {\bibinfo
  {journal} {Acta Crystallographica Section A}\ }\textbf {\bibinfo {volume}
  {49}},\ \bibinfo {pages} {269} (\bibinfo {year} {1993})}\BibitemShut
  {NoStop}%
\bibitem [{\citenamefont {Ji}\ \emph {et~al.}(2023)\citenamefont {Ji},
  \citenamefont {Yu}, \citenamefont {Xu},\ and\ \citenamefont
  {Xiang}}]{PhysRevLett.130.146801}%
  \BibitemOpen
  \bibfield  {author} {\bibinfo {author} {\bibfnamefont {J.}~\bibnamefont
  {Ji}}, \bibinfo {author} {\bibfnamefont {G.}~\bibnamefont {Yu}}, \bibinfo
  {author} {\bibfnamefont {C.}~\bibnamefont {Xu}},\ and\ \bibinfo {author}
  {\bibfnamefont {H.~J.}\ \bibnamefont {Xiang}},\ }\bibfield  {title} {\bibinfo
  {title} {General theory for bilayer stacking ferroelectricity},\ }\href
  {https://doi.org/10.1103/PhysRevLett.130.146801} {\bibfield  {journal}
  {\bibinfo  {journal} {Phys. Rev. Lett.}\ }\textbf {\bibinfo {volume} {130}},\
  \bibinfo {pages} {146801} (\bibinfo {year} {2023})}\BibitemShut {NoStop}%
\bibitem [{\citenamefont {Young}\ and\ \citenamefont
  {Kane}(2015)}]{PhysRevLett.115.126803}%
  \BibitemOpen
  \bibfield  {author} {\bibinfo {author} {\bibfnamefont {S.~M.}\ \bibnamefont
  {Young}}\ and\ \bibinfo {author} {\bibfnamefont {C.~L.}\ \bibnamefont
  {Kane}},\ }\bibfield  {title} {\bibinfo {title} {Dirac semimetals in two
  dimensions},\ }\href {https://doi.org/10.1103/PhysRevLett.115.126803}
  {\bibfield  {journal} {\bibinfo  {journal} {Phys. Rev. Lett.}\ }\textbf
  {\bibinfo {volume} {115}},\ \bibinfo {pages} {126803} (\bibinfo {year}
  {2015})}\BibitemShut {NoStop}%
\bibitem [{\citenamefont {de~la Flor}\ \emph {et~al.}(2021)\citenamefont {de~la
  Flor}, \citenamefont {Souvignier}, \citenamefont {Madariaga},\ and\
  \citenamefont {Aroyo}}]{delaFlor:ug5030}%
  \BibitemOpen
  \bibfield  {author} {\bibinfo {author} {\bibfnamefont {G.}~\bibnamefont
  {de~la Flor}}, \bibinfo {author} {\bibfnamefont {B.}~\bibnamefont
  {Souvignier}}, \bibinfo {author} {\bibfnamefont {G.}~\bibnamefont
  {Madariaga}},\ and\ \bibinfo {author} {\bibfnamefont {M.~I.}\ \bibnamefont
  {Aroyo}},\ }\bibfield  {title} {\bibinfo {title} {{Layer groups:
  Brillouin-zone and crystallographic databases on the Bilbao Crystallographic
  Server}},\ }\href {https://doi.org/10.1107/S205327332100783X} {\bibfield
  {journal} {\bibinfo  {journal} {Acta Crystallographica Section A}\ }\textbf
  {\bibinfo {volume} {77}},\ \bibinfo {pages} {559} (\bibinfo {year}
  {2021})}\BibitemShut {NoStop}%
\bibitem [{\citenamefont {Cornfeld}\ and\ \citenamefont
  {Chapman}(2019)}]{PhysRevB.99.075105}%
  \BibitemOpen
  \bibfield  {author} {\bibinfo {author} {\bibfnamefont {E.}~\bibnamefont
  {Cornfeld}}\ and\ \bibinfo {author} {\bibfnamefont {A.}~\bibnamefont
  {Chapman}},\ }\bibfield  {title} {\bibinfo {title} {Classification of
  crystalline topological insulators and superconductors with point group
  symmetries},\ }\href {https://doi.org/10.1103/PhysRevB.99.075105} {\bibfield
  {journal} {\bibinfo  {journal} {Phys. Rev. B}\ }\textbf {\bibinfo {volume}
  {99}},\ \bibinfo {pages} {075105} (\bibinfo {year} {2019})}\BibitemShut
  {NoStop}%
\bibitem [{\citenamefont {Cornfeld}\ and\ \citenamefont
  {Carmeli}(2021)}]{PhysRevResearch.3.013052}%
  \BibitemOpen
  \bibfield  {author} {\bibinfo {author} {\bibfnamefont {E.}~\bibnamefont
  {Cornfeld}}\ and\ \bibinfo {author} {\bibfnamefont {S.}~\bibnamefont
  {Carmeli}},\ }\bibfield  {title} {\bibinfo {title} {Tenfold topology of
  crystals: Unified classification of crystalline topological insulators and
  superconductors},\ }\href {https://doi.org/10.1103/PhysRevResearch.3.013052}
  {\bibfield  {journal} {\bibinfo  {journal} {Phys. Rev. Res.}\ }\textbf
  {\bibinfo {volume} {3}},\ \bibinfo {pages} {013052} (\bibinfo {year}
  {2021})}\BibitemShut {NoStop}%
\bibitem [{\citenamefont {Park}\ and\ \citenamefont
  {Yang}(2017)}]{PhysRevB.96.125127}%
  \BibitemOpen
  \bibfield  {author} {\bibinfo {author} {\bibfnamefont {S.}~\bibnamefont
  {Park}}\ and\ \bibinfo {author} {\bibfnamefont {B.-J.}\ \bibnamefont
  {Yang}},\ }\bibfield  {title} {\bibinfo {title} {Classification of accidental
  band crossings and emergent semimetals in two-dimensional noncentrosymmetric
  systems},\ }\href {https://doi.org/10.1103/PhysRevB.96.125127} {\bibfield
  {journal} {\bibinfo  {journal} {Phys. Rev. B}\ }\textbf {\bibinfo {volume}
  {96}},\ \bibinfo {pages} {125127} (\bibinfo {year} {2017})}\BibitemShut
  {NoStop}%
\bibitem [{\citenamefont {Lazić}\ \emph {et~al.}(2022)\citenamefont {Lazić},
  \citenamefont {Damljanović},\ and\ \citenamefont
  {Damnjanović}}]{Lazic2022}%
  \BibitemOpen
  \bibfield  {author} {\bibinfo {author} {\bibfnamefont {N.}~\bibnamefont
  {Lazić}}, \bibinfo {author} {\bibfnamefont {V.}~\bibnamefont
  {Damljanović}},\ and\ \bibinfo {author} {\bibfnamefont {M.}~\bibnamefont
  {Damnjanović}},\ }\bibfield  {title} {\bibinfo {title} {Fully linear band
  crossings at high symmetry points in layers: classification and role of
  spin–orbit coupling and time reversal},\ }\href
  {https://doi.org/10.1088/1751-8121/ac7f08} {\bibfield  {journal} {\bibinfo
  {journal} {Journal of Physics A: Mathematical and Theoretical}\ }\textbf
  {\bibinfo {volume} {55}},\ \bibinfo {pages} {325202} (\bibinfo {year}
  {2022})}\BibitemShut {NoStop}%
\bibitem [{\citenamefont {Eick}\ and\ \citenamefont
  {Souvignier}(2005)}]{Eick2005}%
  \BibitemOpen
  \bibfield  {author} {\bibinfo {author} {\bibfnamefont {B.}~\bibnamefont
  {Eick}}\ and\ \bibinfo {author} {\bibfnamefont {B.}~\bibnamefont
  {Souvignier}},\ }\bibfield  {title} {\bibinfo {title} {Algorithms for
  crystallographic groups},\ }\href {https://doi.org/10.1002/qua.20747}
  {\bibfield  {journal} {\bibinfo  {journal} {International Journal of Quantum
  Chemistry}\ }\textbf {\bibinfo {volume} {106}},\ \bibinfo {pages} {316}
  (\bibinfo {year} {2005})}\BibitemShut {NoStop}%
\bibitem [{\citenamefont {Niggli}(1929)}]{doi:10.1080/11035892909447060}%
  \BibitemOpen
  \bibfield  {author} {\bibinfo {author} {\bibfnamefont {P.}~\bibnamefont
  {Niggli}},\ }\bibfield  {title} {\bibinfo {title} {Krystallographische und
  strukturtheoretische grundbegriffe (handbuch der experimentalphysik, bd vii,
  1)},\ }\href {https://doi.org/10.1080/11035892909447060} {\bibfield
  {journal} {\bibinfo  {journal} {Geologiska Föreningen i Stockholm
  Förhandlingar}\ }\textbf {\bibinfo {volume} {51}},\ \bibinfo {pages} {121}
  (\bibinfo {year} {1929})}\BibitemShut {NoStop}%
\bibitem [{\citenamefont {VanLeeuwen}\ \emph {et~al.}(2015)\citenamefont
  {VanLeeuwen}, \citenamefont {Valent{\'\i}n De~Jes{\'{u}}s}, \citenamefont
  {Litvin},\ and\ \citenamefont {Gopalan}}]{VanLeeuwen:dm5060}%
  \BibitemOpen
  \bibfield  {author} {\bibinfo {author} {\bibfnamefont {B.~K.}\ \bibnamefont
  {VanLeeuwen}}, \bibinfo {author} {\bibfnamefont {P.}~\bibnamefont
  {Valent{\'\i}n De~Jes{\'{u}}s}}, \bibinfo {author} {\bibfnamefont {D.~B.}\
  \bibnamefont {Litvin}},\ and\ \bibinfo {author} {\bibfnamefont
  {V.}~\bibnamefont {Gopalan}},\ }\bibfield  {title} {\bibinfo {title} {{The
  affine and Euclidean normalizers of the subperiodic groups}},\ }\href
  {https://doi.org/10.1107/S2053273314024395} {\bibfield  {journal} {\bibinfo
  {journal} {Acta Crystallographica Section A}\ }\textbf {\bibinfo {volume}
  {71}},\ \bibinfo {pages} {150} (\bibinfo {year} {2015})}\BibitemShut
  {NoStop}%
\end{thebibliography}
\end{document}